\documentclass[aps,prd,showkeys,
showpacs,amssymb,eqsecnum,preprintnumbers,nofootinbib,
superscriptaddress]{revtex4}

\usepackage{graphicx}
\usepackage{latexsym}
\usepackage{amsfonts}
\usepackage{bm}

\begin{document}

\begin{figure}[t]
\vspace{-1.4cm}
\hspace{-16.15cm}
\scalebox{0.1}{\includegraphics{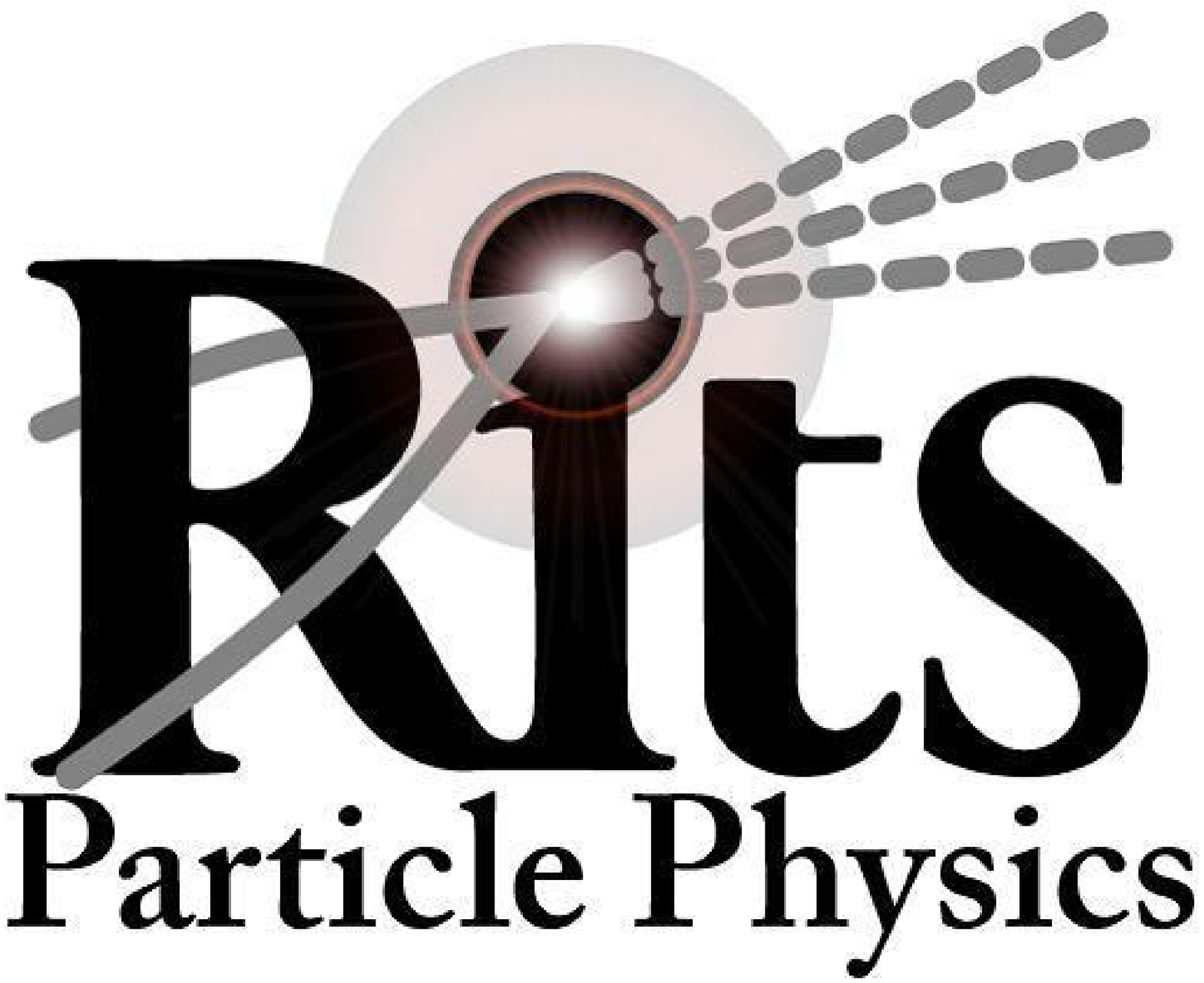}} 
\end{figure}

\newcommand{\vp}{\varphi}
\newcommand{\nn}{\nonumber\\}
\newcommand{\beq}{\begin{equation}}
\newcommand{\eeq}{\end{equation}}
\newcommand{\bed}{\begin{displaymath}}
\newcommand{\eed}{\end{displaymath}}
\def\bea{\begin{eqnarray}}
\def\eea{\end{eqnarray}}

%%%%%%%%%%%%%%%%%%%%%%%%%%%%%%%%%%%%%%%%%%%%%%%%%%%%%%
\title{Quantum fluctuations on a thick de Sitter brane}
\author{Masato~Minamitsuji}
\email[Email: ]{masato@vega.ess.sci.osaka-u.ac.jp}
\affiliation{Department of Earth and Space Science, Graduate School of
Science, Osaka University, Toyonaka 560-0043, Japan}
\affiliation{Yukawa Institute for Theoretical Physics, Kyoto University, 
Kyoto 606-8502, Japan}
\author{Wade~Naylor}\email[Email: ]{naylor@se.ritsumei.ac.jp}
\affiliation{Department of Physics, Ritsumeikan University, 
Kusatsu, Shiga 525-8577, Japan}
\author{Misao~Sasaki}
\email[Email: ]{misao@yukawa.kyoto-u.ac.jp}
\affiliation{Yukawa Institute for Theoretical Physics, Kyoto University, 
Kyoto 606-8502, Japan}

\begin{abstract}
We investigate quantum fluctuations on a de Sitter (dS) brane,
which has its own thickness, in order to examine whether or not the finite
thickness of the brane can act as a {\it natural}  cut-off for the 
Kaluza-Klein (KK) spectrum. 
We calculate the amplitude of the KK modes and the bound state by using
the zeta function method after a dimensional reduction.
We show that the KK amplitude is finite for a given brane
thickness and in the thin wall limit the standard surface divergent 
behavior is recovered.
The strength of the divergence in the thin wall limit depends on the number 
of
dimensions, e.g., logarithmic on a two dimensional brane
and quadratic on a four dimensional brane.
We also find that the amplitude of the bound state mode and KK modes 
depends on the choice of renormalization scale; and for fixed renormalization 
scales the bound state mode 
is insensitive to the brane thickness both for two and
four-dimensional dS branes. 
\end{abstract} 

\pacs{04.50.+h; 98.80.Cq}
\keywords{Extra dimensions, Zeta function regularization}
\preprint{YITP-05-40}
\preprint{RITS-PP-001}
\date{\today}
\maketitle

%%%%%%%%%%%%%%%%%%%%%%%%%%%%%%%%%%%%%%%%%%%%%%%%%%%%%%%%%%%%%%%%%%%%%
\section{Introduction}

Recent progress in string theory suggests that our 
universe is not four-dimensional, in reality, but is a
four-dimensional submanifold, called a ``{\it brane}'' embedded 
into a higher dimensional spacetime, called the ``{\it bulk}''.
The model which was proposed by Randall and Sundrum (RS)
succeeds in the localization of gravity around the brane due to
 a fascinating process, i.e., through the warping of the
extra-dimension \cite{Randall:1999vf}. 
This model has been given phenomenological grounds from various
aspects of higher-dimensional theories of gravity, e.g.,
in terms of cosmology, higher-dimensional black holes and the AdS/CFT
correspondence~(see e.g.,~\cite{Maartens:2003tw, Brax:2004xh} for reviews).

In spite of much effort by various authors, there seem to be few
theoretical predictions of such a braneworld with cosmological 
observations.
In the RS type brane model, information from a
higher-dimensional gravity theory is carried by~Kaluza-Klein
(KK) modes, which are massive modes from the observers' point of view, who are 
living on the brane. Thus, if KK modes are detected in future
observations this may uniquely determine whether or not our universe
is a brane. 
 
The quantum fluctuations that are usually assumed to be produced
during the inflationary phase are considered as natural seeds for 
the perturbations, which form the various cosmological structures and
produce the CMB anisotropy we see today.
Thus, in the braneworld scenario, the quantum fluctuations may imprint
information from the KK modes on today's cosmological observations.   

In some braneworld models, the inflaton whose dynamics induces
inflation on the brane is naturally set into the bulk as a result of
a dimensional reduction of some higher-dimensional gravitational theory,
e.g., dilaton or moduli fields, 
as discussed in \cite{Kobayashi:2000yh,Himemoto:2000nd,Koyama:2003yz}. 
In various bulk inflaton models, quantum fluctuations have been well
examined~\cite{Kobayashi:2000yh,Koyama:2003yz,Sago:2001gi,Naylor:2004ua,Pujolas:2005mu}.
However, it is well-known that a Casimir surface
divergence remains on the brane even after UV regularization. This type of 
divergence prevents us from evaluating quantum fluctuations exactly on the 
brane, though there are approaches to deal with such a problem, e.g., see 
\cite{Naylor:2004ua, Pujolas:2005mu}. In all previous cases mentioned above it was 
assumed that the brane was infinitesimally thin,
which is inherited from the original proposal by RS. 

{}From a more realistic point of view; however, it is rather natural for the 
brane to have a thickness.
Furthermore, string theory, which brane world is based upon, has a
minimum length scale, i.e., the string length scale
$l_s=\sqrt{\alpha'}$, where $\alpha'$ is the inverse of string tension.  
This supports the possibility that, in reality, the brane has a
finite thickness, rather than being infinitesimally thin. 
Various thick brane models have been discussed, see e.g.,~\cite{DeWolfe:1999cp,%
Gremm:1999pj,Csaki:2000fc,Arefeva:2000qv,Giovannini:2001xg,Kobayashi:2001jd,%
Sasakura:2002tq,Sasakura:2002hv,Ghoroku:2003kc,Bronnikov:2003gg}.
Then, we are interested in whether or not ``{\it  the finite thickness
of the brane, if it exists, acts as a natural cut-off for the KK
spectrum.''} In this article, we discuss this speculative idea by
investigating the behavior of the KK modes in an explicit thick de
Sitter (dS) brane model.
 
We consider a simple thick dS brane model which is supported by a bulk
scalar field $\phi$ with an axion-like potential, originally discussed by
Wang~\cite{Wang:2002pk, Guerrero:2005xx}.
Then, we introduce another scalar field $\chi$  
as a bulk inflaton \cite{Kobayashi:2000yh, Himemoto:2000nd} type field  
and investigate its quantum fluctuations.
We use the zeta function method in conjunction with the dimensional
reduction approach developed in \cite{Naylor:2004ua}.
We show that the brane thickness regularizes the UV behavior of the
KK modes and compare its amplitude with that of the bound
state mode, which is relevant from the observational point of view.

This article is organized as follows: in Sec.~II, we introduce a thick dS 
brane model supported by a bulk scalar field.
In Sec.~III, bearing the bulk inflaton model in mind, we introduce
another quantized scalar field and make the necessary preparations to 
evaluate
the amplitude. We take the dimensional reduction approach and
calculate the amplitude using generalized zeta functions. 
In Sec.~IV, focusing on a two dimensional dS brane (three-dimensional
bulk),
we calculate the amplitude of quantum fluctuations and show
that the brane thickness acts as a natural cut-off for the KK
spectrum.
The surface divergence in the thin wall limit is logarithmic.
We also show that the bound state amplitude depends on the
renormalization scale; however, for a fixed renormalization scale it is 
almost
independent of the brane thickness.
Then in Sec.~V, we discuss quantum fluctuations on a four
dimensional dS brane (five-dimensional bulk) and obtain results
similar to that for the two-dimensional case.
The leading order surface divergence in the thin wall limit is
quadratic in this case.
In Sec.~VI, we summarize our results and mention future work
related to these issues.
In Appendix A, we derive the fluctuation amplitude of the bound state
on two and four dimensional dS branes, respectively.
In Appendix B, we demonstrate the classical stability of the thick brane
model both against tensor and scalar perturbations.

%%%%%%%%%%%%%%%%%%%%%%%%%%%%%%%%%%%%%%%%%%%%%%%%%%%%%%%%%%%

\section{A thick de Sitter brane model}

We consider the Einstein theory coupled to a bulk scalar field,
\begin{eqnarray}
S=\frac{1}{2}\int d^{d+1}x \sqrt{-g}
      \Bigl(            \stackrel{(d+1)}{R}
           -\bigl(\partial \phi\bigr)^2
           -2V(\phi)
      \Bigr)\,,\label{action1}
\end{eqnarray}
where the potential of the scalar field is given by the axion like form~\cite{Wang:2002pk, Guerrero:2005xx},
\begin{eqnarray}
V(\phi)= V_0 \Bigl( \cos\Bigl[\frac{\phi}{\phi_0}\Bigr]
                       \Bigr)^{2(1-\sigma)}\,.
\end{eqnarray}
Note that we shall set $\kappa_{d+1}^2=1$ in this article.

We shall assume a static configuration, namely $\phi$ depends on only the
bulk coordinate and make the following metric ansatz
\begin{eqnarray}
ds^2= b(z)^2(dz^2+\gamma_{\mu\nu}dx^{\mu}dx^{\nu})\,,
\end{eqnarray}
where $\gamma_{\mu\nu}$ denotes the metric of $d$-dimensional de
Sitter~(dS) spacetime.
Following the above ansatz, we obtain the Einstein equations
\begin{eqnarray}   
   && \frac{d(d-1)}{2}\Bigl(\frac{b'}{b}\Bigr)^2 
      -\frac{d(d-1)}{2}H^2
     =\frac{1}{2}\phi'^2 -b^2V\,,
\nonumber \\
&& (d-1)\Bigl[
     \frac{b''}{b} 
      +\frac{1}{2}(d-4)\Bigl(\frac{b'}{b}\Bigr)^2
      -\frac{1}{2}(d-2)H^2\Bigr]
     =-\frac{1}{2}\phi'^2 -b^2V
\end{eqnarray}
and the field equation for the scalar field is
\begin{eqnarray}
 \phi'' + (d-1) \frac{b'}{b} \phi' -b^2 \frac{\partial V}{\partial 
\phi}=0\,,
\end{eqnarray}
where the prime $\,'\,$ denotes the derivative with respect to $z$.
Note that only two of these three equations are independent.
For this potential, we find the solutions
\begin{eqnarray}
  b(z)=\left(\cosh\left(\frac{H z}{\sigma}\right)\right)^{-\sigma}\,,\qquad
 \phi(z)= \phi_0 \sin^{-1}\left(\tanh\left(\frac{H z}{\sigma} \right)\right)\,,
\label{thickbranesolution}
\end{eqnarray}
where
\begin{eqnarray}
H^2 = \frac{2\sigma V_0}{(d-1)[1+(d-1)\sigma]} \,,
 \quad
\phi_0 = \sqrt{(d-1)\sigma(1-\sigma)}~.
\end{eqnarray}
This solution represents a dS domain wall whose energy is
localized at $z=0$,~i.e.,~the center of the wall.
The parameter $\sigma$ has the meaning of the thickness of the wall
(brane) from the physical point of view. 
In order to keep the positivity in the square root, we should
restrict the range to \cite{Wang:2002pk}
\beq
0< \sigma< 1~.
\eeq

%%%%%%%%%%%%%%%%%%%%%%%%%%%%%%%%%%%%%%%%%%%%%%%%%%%%%%%%%%%%%%%%%%%%%%%%%%%
\section{Quantized scalar field perturbations}

Our purpose is to discuss the quantized scalar field perturbations 
on a thick, inflating brane model. We achieve this by introducing
another scalar field $\chi$, which is coupled to the domain wall
configuration and its fluctuations. Hence, we add the action of the
scalar field $\chi$ to the original action Eq. (\ref{action1}), i.e.,
\begin{eqnarray}
S=\frac{1}{2}\int d^{d+1}x \sqrt{-g}
      \Bigl(
            \stackrel{(d+1)}{R}
           -\bigl(\partial \phi\bigr)^2
           -2  V(\phi)
      \Bigr)
+
      \frac{1}{2}\int d^{d+1}x \sqrt{-g}
      \Bigl(
           -\bigl(\partial \chi\bigr)^2
           -\xi \stackrel{(d+1)}{R} \chi^2
      \Bigr)\,,
\end{eqnarray}
where $\xi$ is the scalar curvature coupling.

As discussed in \cite{Olum:2002ra, Graham:2002yr} the coupling of the
field $\chi$ to $\phi$
can be ignored because its backreaction to the domain wall geometry is
only important at higher order, $O(\chi^2)$.\footnote{These works 
use~the methods developed in \cite{Graham:2002xq}.}
This assumption allows us to treat the $\chi$-field contribution
perturbatively.~The minimally coupled
case, $\xi=0$,~will be of particular interest,
because the perturbation equations are very similar
to those for tensor perturbations of the metric
(see Appendix B).

%%%%%%%%%%%%%%%%%%%%%%%%%%%%%%%%%%%%%%%%%%%%%%%%%%%%%%%%%%%%%%%%%%%
\subsection{Dimensional reduction approach}

We shall evaluate the amplitude of the quantum field $\chi$ based on a
dimensional reduction of the higher dimensional canonically quantized
field. This method has been already discussed in \cite{Naylor:2004ua}
and we refer the reader to this reference for more details.

In this method, the action of $\chi$ is rewritten as
\bea
 S_{\chi}
=     \frac{1}{2}\int d^{d+1}x \sqrt{-g}\,
  \chi    \Bigl(
            \Box_{d+1}
           -\xi \stackrel{(d+1)}{R}
      \Bigr)\chi\,,
\eea
where we set a regulator boundary at $z=L$ in order to obtain a {\it
well-posed} quantum field theory on the dimensionally reduced spacetime.
Then, the bulk modes become discrete
and the solution is written as 
\bea
\chi(z,x^\mu)=\sum_n F_n(z)\varphi_{n}(x^{\mu}) H^{1/2}\,,
\eea
where $\varphi_n$ has the dimension of a scalar field in the
$d$-dimensional dS
spacetime.~Due to the maximal symmetry of dS spacetime,~we can
integrate out
the dependence on the transverse directions, $x^{\mu}$,
assuming that the vacuum respects the dS invariance.
Hence, we shall drop it in the amplitude.

Integrating the action with respect to $z$,~it is reduced to the
summation of theories of a 
$d$-dimensional massive scalar field with mass $m_n$: 
\bea
S_{\chi}
=     \frac{1}{2}\sum_{n}\int d^{d}x \sqrt{-\gamma}\,
  \varphi_n(x^{\mu})    \Bigl(
            \Box_{d}
           -m_n^2
      \Bigr)
  \varphi_n(x^{\mu})\,,
\eea 
where we employed the normalization condition
\bea
2\int^L_0 (H \, dz) \,b^{d-1}(z)
  F_{q_n}(z) F_{q_n'}(z)
= \delta_{n n'}\,.
\eea
Note that the multiplying factor of two is due to the 
${\mathbb Z}_2$-symmetry.
%Thus, the $(d+1)$-dimensional theory is reduced to a linear combination of 
%lower, $d$-dimensional scalar field
%theories with masses
The mass-squared $m_n^2$ is given by
\bea
m_n^2=q_n^2H^2+\frac{(d-1)^2}{4}H^2\,.
\eea 

We introduce a new function $f_{q_n}(z):= b^{(d-1)/2}(z) F_{q_n}(z)$, which
obeys the Schr{\"o}dinger like equation
\bea
-f_{q_n}''(z) 
+\tilde V(z) H^2 f_{q_n}
= q_n^2H^2 f_{q_n}(z)\,,
\eea
where
\bea
\tilde V(z):=
  -\bigl(\xi_c -\xi\bigr)
  \Bigl\{
     d(d-1)
   +\frac{2d}{\sigma}
  \Bigr\}
\frac{1}{\cosh^2(H z/\sigma)}\,.
\label{fieldpotential}
\eea
For the minimally coupled case,~$\xi=0$,~this potential reduces to the one 
for %tensor%%%%%%%%%%%%%%%%%%%%%%%%
the tensor metric perturbations (i.e.,~gravitons), Eq. (\ref{tensorpotential}).
The reason that we choose a massless scalar field as a probe in this article is
to obtain some insight into the graviton case (and also for technical simplicity). 
In Fig.~1, we plot the potential for the $d=4$ case explicitly for~$\xi=0$. 
It is evident that the potential becomes deeper for smaller values 
of $\sigma$. 
%%%%%%%%%%% Figure  1 %%%%%%%%%%%%%%%%%%%%%%%%%%%%%%%%%%%
\begin{figure}
\begin{center}
  \begin{minipage}[t]{.45\textwidth}
   \begin{center}
    \includegraphics[scale=.85]{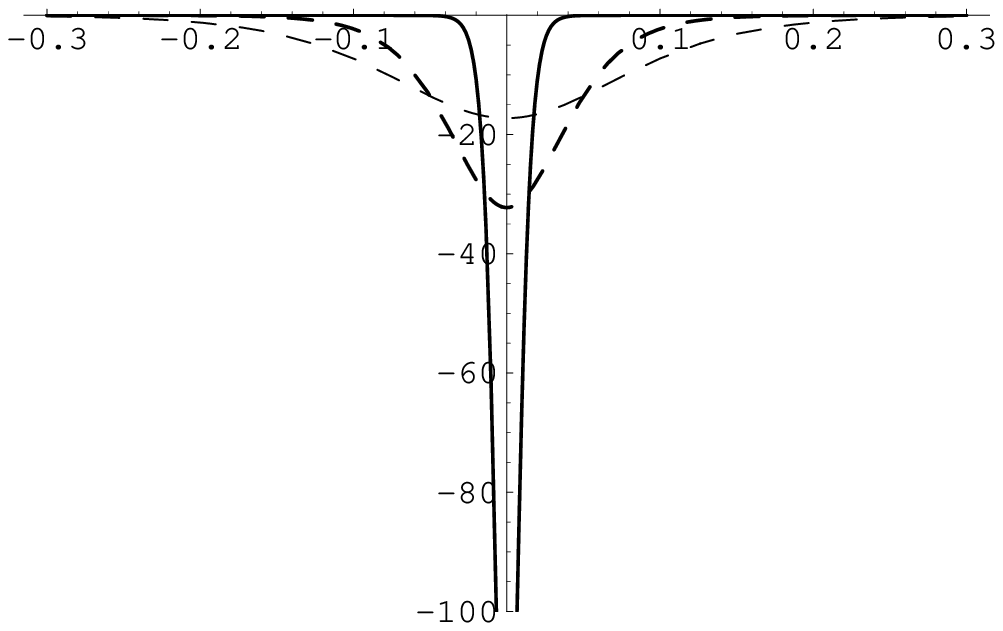}
        \caption{
        The potential for a minimally coupled bulk scalar field perturbation is shown as a function of $Hz$ for a four-dimensional dS wall.
        The thick, thick-dashed and dashed curves correspond to the cases 
of
        $\sigma=0.01, 0.05, 0.1$, respectively.}  
   \end{center}
   \end{minipage} 
\hspace{0.5cm}
   \begin{minipage}[t]{.45\textwidth}
   \begin{center}
    \includegraphics[scale=.85]{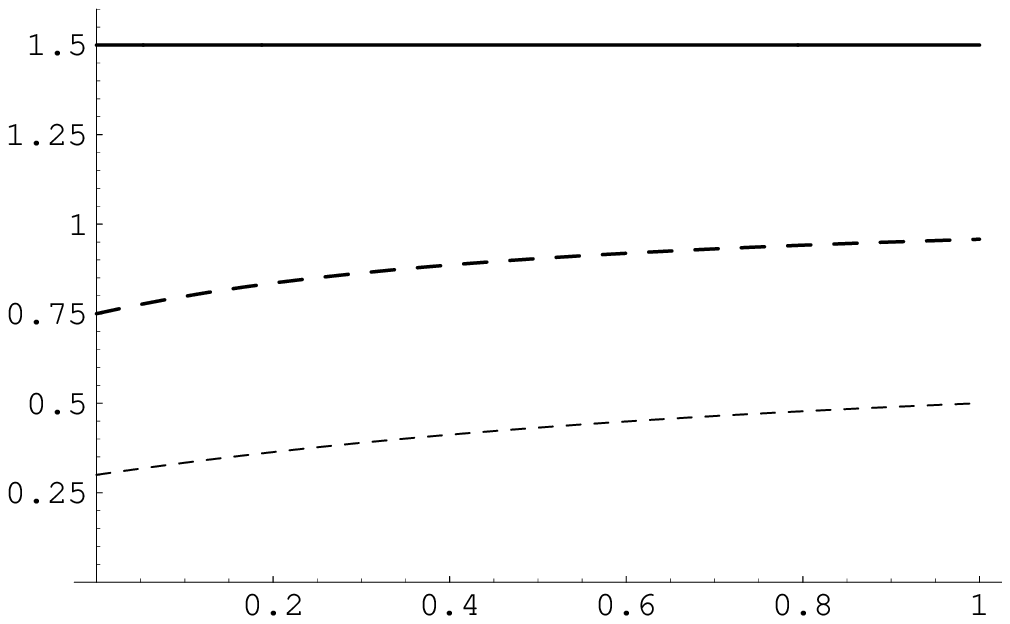}
\caption{
$\nu/\sigma$ is plotted as a function of thickness, $\sigma$.
The thick, thick-dashed and dashed curves correspond to the cases 
for $\xi=0, 3/32, 3/20$, respectively.}          
   \end{center}
   \end{minipage}
   \end{center}
\end{figure}
%%%%%%%%%%%%%%%%%%%%%%%%%%%%%%%%%%%%%%%%%%%%%%%%%%%%%

The solution of the KK modes is
\begin{eqnarray}
f_{q_n}(z)= C_1 P^{i\sigma q_n}_{\nu}(x)
    + C_2 P^{-i\sigma q_n}_{\nu}(x)\,,
\end{eqnarray}
where $P^{\mu}_{\nu}(x)$ denotes the Legendre functions of the first kind,
$x:=\tanh(H z/\sigma)$ and
\begin{eqnarray} 
\nu:= \frac{  \sqrt{1+4 (\xi_c-\xi)
               ( d(d-1)\sigma^2+2d\sigma)}
                       -1}{2}\,.\label{nudef}
\end{eqnarray}
The coupling
\bea
\xi_c =\frac{d-1}{4d}
\eea
denotes the  conformal coupling strength, e.g., for the $d=2$ case 
$\xi_c=1/8$
and for the $d=4$ case $\xi_c=3/16$.
In this article, we restrict the coupling to the range $0\leq \xi
\leq \xi_c$.

The mass of the bound state mode is given by
\begin{eqnarray} 
 q_0
 =\frac{i\nu}{\sigma}\,,
\end{eqnarray}
which has a maximal value of $(d-1)/2$ at $\xi=0$. 
For $\xi<0$, the bound state becomes tachyonic
and non-normalizable. For $\xi=0$, it is the zero mode and for $\xi=\xi_c$
it becomes the lowest mass KK mode, irrespective of the choice of $\sigma$.
In Fig.~2, we plot $\nu/\sigma$ as a function of $\sigma$ for several
choice of $\xi$.
We find that this ratio is almost independent of $\sigma$. 
Note that the $\xi=0$ case is equivalent to the bulk tensor perturbations.

%%%%%%%%%%%%%%%%%%%%%%%%%%%%%%%%%%%%%%%%%%%%%%%%%%%%%%%%%%%%%%%%%%%%%%
\subsection{The zeta function method}

Given the functions $f_{q_n}(z)$,~the vacuum expectation value is defined by
\bea
 \langle \chi^2(z) \rangle
=\frac{2H}{b^{d-1}(z)}
\sum_{n}  f_{q_n}^2(z) 
         \langle
            \varphi_{q_n}^2(x)
         \rangle\,,\label{sum}
\eea
where the factor of two is due to the ${\mathbb Z}_2$-symmetry.
{}From now on, we shall discuss the quantized field theory in Euclidean
space, i.e., the metric is
\bea
ds^2=b^2(z)\bigl(
            dz^2 + d\Omega_d^2
           \bigr)\,,
\eea
where $d\Omega_d$ is the line element of $S^{d}$ with unit radius,
whose volume is given by
\bea
V_{S^{d}}=
\frac{2\pi^{\frac{d+1}{2}}}{\Gamma(\frac{d+1}{2})}\,.
\eea
Thus, in order to consider the quantum fluctuations of a $d$-dimensional
field, we assume that the vacuum is given by the Euclidean vacuum,
which corresponds to the dS invariant, Bunch-Davis vacuum in the
original Lorentzian spacetime.

For the $d$-sphere, $S^d$, any local quantities are related to global ones 
by
simply dividing by the volume of the sphere (a property of maximally
symmetric spaces; see \cite{Naylor:2004ua}).  
Thus, we are particularly interested in the local vacuum expectation
value as only a function of $z$ (one non-trivial dimension), implying
\bea
K_n(t)
=\int d^d{x} \sqrt{\gamma}K_n(x,x;t)\,,
\eea
where $K_n$ is the dS heat kernel for each mode $n$, see
\cite{Naylor:2004ua}. 
Thus, due to the maximal symmetry of dS space, the global heat
kernel is simply related to the local one by 
\bea
K_n(t)
=\frac{V_{S^d}}
      {H^{d}}
   K_n(x,x;t)~.
\eea 

At this stage it is convenient to rescale the amplitude as
\bea
  \langle \tilde \chi^2(z) \rangle
=  \frac{b(z)^{d-1} V_{S^d}}{H^{d+1}} \langle \chi^2(z) \rangle\,,
\eea
where overall factors can be restored at the end of the
calculation. Now we may sum up all the KK modes in Eq. (\ref{sum});
however, as is well known, a naive summation over all the KK modes gives
rise to unwanted divergences.

To deal with such a problem, we construct the {\it local} zeta
function, $\zeta(z,s)$, along the lines of reference
\cite{Naylor:2004ua}, where the parameter $s$ is 
initially assumed to be $s>(d+1)/2$ in $(d+1)$-dimensions. 
Once we obtain such a zeta function, after analytic continuation to $s\to
1$, we end up with
\begin{eqnarray}
\langle \tilde \chi^2(z) \rangle
=\lim_{s\to 1}\tilde \zeta(z,s)\,,
\end{eqnarray}  
where
\begin{eqnarray}
 \tilde{\zeta}(z,s)
:= \frac{b(z)^{d-1} V_{S^d}}{H^{d+1}}
   \zeta(z,s)
=\frac{1}{\Gamma(s)}
\int^{\infty}_0 dt\, t^{s-1}K(z,t)\,.
\end{eqnarray}
$K(z,t)$ is the local heat kernel defined as
\begin{eqnarray}
K(z,t)=2 \sum_{n=1}^{\infty} 
          f_{q_n}^2(z) K_n(t) \,,
\end{eqnarray}
and
\begin{eqnarray}
 K_{n}(t)
=\sum_{j=0}^{\infty} d_j\,e^{-[q_n^2+(j+1/2)^2]H^2 t}\,,
\end{eqnarray}
where $d_j$ is the degeneracy for each mode $j$ given by
\bea
d_j=\left(2j+d-1\right)\frac{(j+d-2)!}{j!(d-1)!}\,,
\eea
is the global heat kernel for each KK mode.
%We shall first concentrate on the case $d=2$, i.e., the degeneracy 
%$d_j=2j+1$. 
Note that the dimension of $\tilde{\zeta}$ is slightly different
to the case discussed in \cite{Naylor:2004ua}, because of a difference
in dimension of the warp factor.

%%%%%%%%%%%%%%%%%%%%%%%%%%%%%%%%%%%%%%%%%%%%%%%%%%%%%%%%%%%%%%%%%%%

\subsection{Contour integral representation of the local zeta
function}

First, as a resolution to the {\it subtle} nature of the continuous
modes, we introduce {\it another boundary} at $z=L$. This then enables
us to evaluate the zeta function using the residue theorem, based on
certain assumptions relating to the zeros of the function in the
contour. Then, after constructing such a zeta function we show that we
can take the one brane limit $L\to\infty$ in a well defined
manner.\footnote{The same approach cannot be used for the one-loop
effective action because it is a global quantity, e.g., see the
discussion in \cite{Pujolas:2005mu}.} 
 
The solution for the scalar field perturbations in general
dimensions is given by
\begin{eqnarray}
&& f_q(z)
=N_q \Bigl(\alpha_q  P^{iq\sigma}_{\nu}(x)
-\beta_q R^{iq\sigma}_{\nu}(x)\Bigr)\,,
\end{eqnarray}
where for convenience, we choose the second solution 
$R^{iq\sigma}_{\nu}(x)$
to satisfy
\begin{eqnarray}
P^{iq \sigma }_{\nu}(x) R^{iq \sigma }_{\nu}{}'(x)
-R^{iq \sigma }_{\nu}(x) P^{iq \sigma }_{\nu}{}'(x)
=\frac{1}{1-x^2}\,,\label{Wronski}
\end{eqnarray}
where $x=\tanh(Hz/\sigma)$.
There are several candidates for $R^{iq\sigma}_{\nu}(x)$ such as
\begin{eqnarray}
\frac{\Gamma(-iq\sigma+\nu+1)}
     {\Gamma(iq\sigma+\nu+1)}
     Q^{iq\sigma}_{\nu}(x)\,,
\quad
-\frac{\pi}{2i\sinh(\pi q\sigma)} P^{-iq\sigma}_{\nu}(x)\,,
\end{eqnarray}
and so on.
For now we do not need to specify the explicit form of the second
solution $R^{iq\sigma }_{\nu}(x)$, but only use the property of the
Wronskian in Eq. (\ref{Wronski}).  

To be specific, let us consider the case of Neumann boundary conditions.
The boundary conditions at the center of the thick brane and the second
boundary are respectively
\begin{eqnarray}
&&f_q'(z)|_{z=0}=0\,,\quad
f_q'(z)|_{z=L}=0\,.
\end{eqnarray}
Note, the thick brane is not a boundary, we just fix
 the $z$~derivative of the field at a point to 
obtain a well-posed eigenvalue equation. From these boundary conditions, we get
an equation which determines the KK mass spectrum as 
\begin{eqnarray}
&& P^{iq\sigma}_{\nu}{}'(0)
   R^{iq\sigma}_{\nu}{}'(x_L)
 - P^{iq\sigma}_{\nu}{}'(x_L)
   R^{iq\sigma}_{\nu}{}'(0)
 =0\,.
\end{eqnarray}
We denote the solutions for the eigen-equation as
$q_n~(n=1,2,3,\cdots)$ whose eigenfunctions are 
\begin{eqnarray}
f_{q_n}(z)=N_{q_n} 
    \Bigl(
      \alpha_{q_n}P^{iq_n\sigma}_{\nu}(x)
   - \beta_{q_n}R^{iq_n\sigma}_{\nu}(x)
    \Bigr)\,,
\end{eqnarray}
where
\begin{eqnarray}
 \frac{\alpha_{q_n} }{\beta_{q_n}}
&=&
\frac{R^{iq_n\sigma}_{\nu}{}'(0)}
      {P^{iq_n\sigma}_{\nu}{}'(0)}
=\frac{R^{iq_n\sigma}_{\nu}{}'(x_L)}
      {P^{iq_n\sigma}_{\nu}{}'(x_L)}\,.
\end{eqnarray}
We assume $q_1<q_2<q_3<\cdots$, respectively. 
Note that the final equality is satisfied only for $q=q_n$.  
Without loss of generality, we can choose $\alpha_q= R^{iq\sigma}_{\nu}{}'(0)$ and $\beta_q = P^{iq\sigma}_{\nu}{}'(0)$.
We shall also require the normalization constant for $n$-th mode which
is found to be  
\begin{eqnarray}
 N^{-2}_{q_n}&=&2\int^L_0
   (H\,dz)
   \Bigl(
   \alpha_{q_n} P^{iq_n\sigma}_{\nu}(x)
  - \beta_{q_n} R^{iq_n\sigma}_{\nu}(x)
   \Bigr)^2
\nonumber \\
&=& -\frac{1}{\sigma q_n} 
    \frac{R^{iq_n\sigma}_{\nu}{}'(0)}
        {R^{iq_n\sigma}_{\nu}{}'(x_L)} 
  \partial_{q}
   \Bigl(
 %  \Bigl(
%    \alpha_{q}
    R^{iq\sigma}_{\nu}{}'(0)
    P^{iq\sigma}_{\nu}{}'(x_L)
  -% \beta_{q}
    P^{iq\sigma}_{\nu}{}'(0)
    R^{iq\sigma}_{\nu}{}'(x_L)
 % \Bigr)
 %-\Bigl(
 %  \alpha_{q} P^{iq\sigma}_{\nu}{}'(0)
 % - \beta_{q} R^{iq\sigma}_{\nu}{}'(0)
 %  \Bigr)
   \Bigr)_{q=q_n}\,.
\end{eqnarray}
Note that in the second step we used Eq. (3.8) and
integrated by parts.

Now we have all the necessary tools to calculate the zeta function by
applying the residue theorem as follows: from the equations given
above, the normalized mode functions can be written as
\begin{eqnarray}
&&f_{q_n}^2(z)=
     \frac{\sigma q_n G(q_n,z)}{ \partial_q F(q)|_{q=q_n}}\,,
\end{eqnarray}
where, 
\begin{eqnarray}
&& F(q)=-\Bigl(
            R^{iq\sigma}_{\nu}{}'(0)
            P^{iq\sigma}_{\nu}{}'(x_L)
          - P^{iq\sigma}_{\nu}{}'(0) 
            R^{iq\sigma}_{\nu}{}'(x_L) 
          \Bigr)
%          -\Bigl(
%           \alpha_{q_n} P^{iq_n\sigma}_{\nu}{}'(x_L)
%          - \beta_{q_n} R^{iq_n\sigma}_{\nu}{}'(x_L) 
%           \Bigr)
%          +\Bigl(
%            \alpha_{q_n} P^{iq_n\sigma}_{\nu}{}'(0)
%          - \beta_{q_n} R^{iq_n\sigma}_{\nu}{}'(0) 
%           \Bigr)\,,
\end{eqnarray}
and
\beq
G(q,z)=\Bigl(
           R^{iq\sigma}_{\nu}{}'(0)
           P^{iq\sigma}_{\nu}(x)
          -P^{iq\sigma}_{\nu}{}'(0)
           R^{iq\sigma}_{\nu}(x)
       \Bigr)
       \Bigl(
           R^{iq\sigma}_{\nu}{}'(x_L)
           P^{iq\sigma}_{\nu}(x)
          -P^{iq\sigma}_{\nu}{}'(x_L)
           R^{iq\sigma}_{\nu}(x)
       \Bigr)\,.
  %   \alpha_{q_n}^2 R^{iq_n\sigma}_{\nu}(x)^2
  %    -2\alpha_{q_n}\beta_{q_n}
  %   P^{iq_n\sigma}_{\nu}(x)
  %   R^{iq_n\sigma}_{\nu}(x)
  % + \beta_{q_n}^2 R^{iq_n\sigma}_{\nu}(x)^2
%\Bigr)\,.
\eeq
This form is essential in order to apply the residue theorem. 
%%%%%%%%%%%%%%%%%%%%%%%%%%%%%%%%%%%%%%%%%%%%
Whence, the zeta function can be written as a
contour integral in the complex $u$ plane 
\begin{eqnarray}
\tilde{\zeta}(z,s)
&=&2\mu^{2(s-1)}
    \sum_{n=1}^{\infty}\,\sum_{j=0}^{\infty}
    \frac{d_j \,f_{q_n}^2(z)}
        {\bigl[q_n^2+(j+{d-1\over 2})^2  \bigr]^sH^{2s}}
\nonumber 
\\
&=&
2  \mu^{2(s-1)}
  \sum_{n=1}^{\infty}\sum_{j=0}^{\infty}
   \frac{\sigma  q_n G(q_n,z)}{ \partial_q F(q)|_{q=q_n} }
{d_j\over \bigl[q_n^2+(j+{d-1\over 2})^2  \bigr]^sH^{2s}}
\nonumber \\
&=&
2\mu^{2(s-1)}
  \oint_{C} \frac{du}{2\pi i}
 \frac{\sigma  u G(u,z)}{  F(u)|_{u=q_n} } 
\sum_{j=0}^{\infty}\frac{d_j}
               {[u^2+(j+{d-1\over 2})^2]^sH^{2s}}\,\,,
\label{int_closed}
\end{eqnarray}
where the poles at $u=q_n$ are on the positive side of real axis and
therefore, the closed contour $C$ has to be taken around the positive 
real axis in general.
Note that we have introduced a mass scale $\mu$ to
keep the dimension. This term is in fact the renormalization scale and 
groups
with any divergent terms in the expression for the amplitude. 
Then, given the fact that there are no
poles in the complex $u$-plane, besides those on the real axis, 
we can naturally deform the contour $C$ into $C'$ (see Fig.~3)
\bea
 \tilde{\zeta}(z,s)
=2\mu^{2(s-1)}
  \oint_{C'} \frac{du}{2\pi i}
 \frac{\sigma  u G(u,z)}{  F(u)|_{u=q_n} } 
\sum_{j=0}^{\infty}\frac{d_j}
               {[u^2+(j+{d-1\over 2})^2]^sH^{2s}}\,,
\eea
which is composed of a
line parallel to the
imaginary axis with a small real part and a large semi circle
on the positive real half of the complex plane, which is depicted 
in Fig 3. As we mentioned previously,
initially keeping $s$ larger than $(d+1)/2$, the contribution from
the larger semi-circle becomes negligible. 

%%%%%%%%%%% Figure  Contour %%%%%%%%%%%%%%%%%%%%%%%%%%%%%%%%%%%
\begin{figure}
\begin{center}
  \begin{minipage}[t]{.45\textwidth}
   \begin{center}
    \includegraphics[scale=.55]{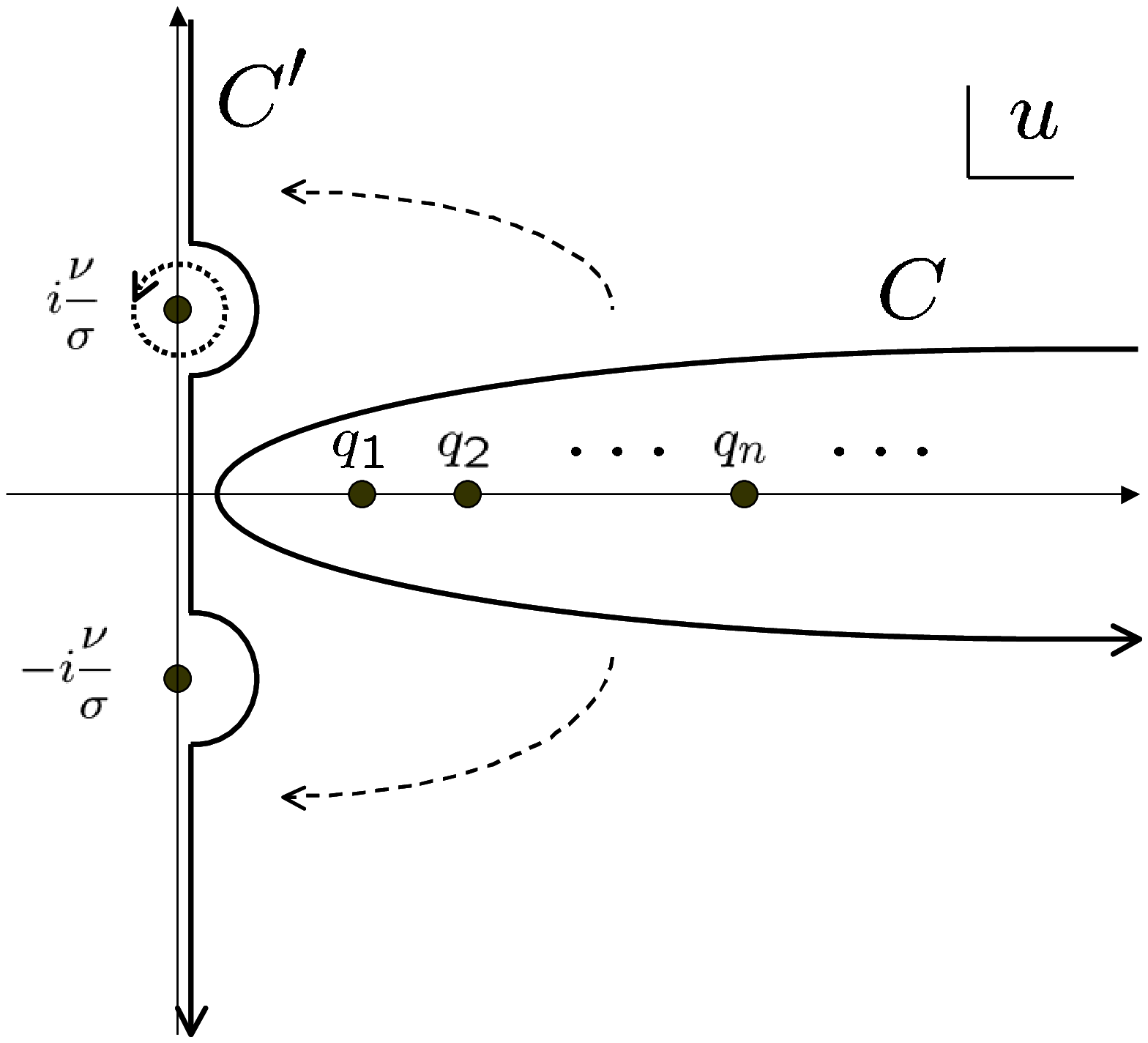}
        \caption{
The contour $C$, used to evaluate the KK amplitude.
The poles on the real axis $q_i\, (i=1,2,\cdots)$ correspond to the KK
modes, while those on the imaginary axis correspond to the bound state.
We can deform $C$ into $C'$ because there are no poles 
in the complex plane besides those on the real and imaginary axes.
The closed contour depicted by the dotted line is used to evaluate the
bound state amplitude.}
   \end{center}
   \end{minipage} 
\hspace{0.5cm}
   \begin{minipage}[t]{.45\textwidth}
   \begin{center}
    \includegraphics[scale=.55]{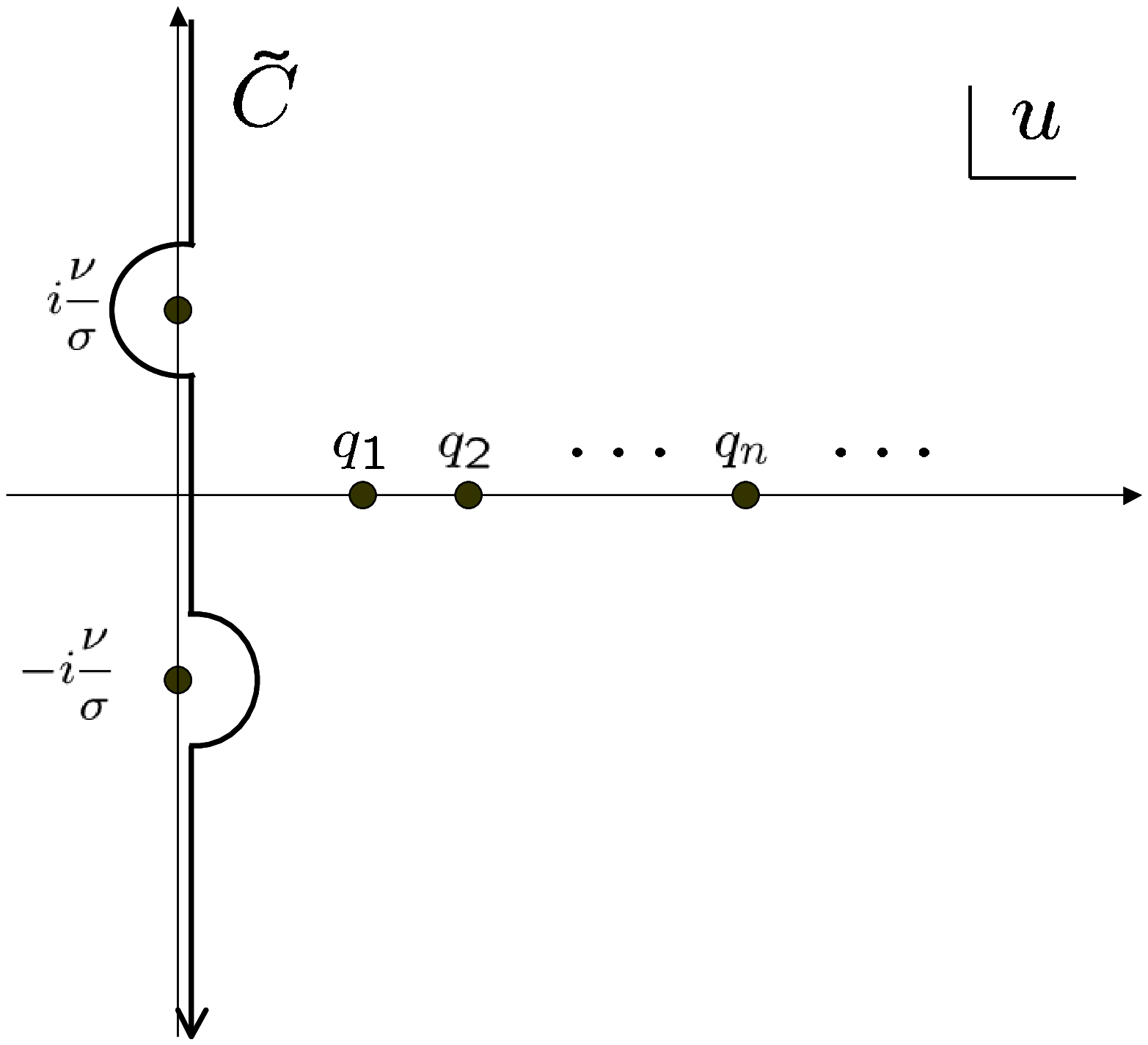}
\caption{
The contour $\tilde C$, used to  evaluate the total amplitude. 
}
   \end{center}
   \end{minipage}
   \end{center}
\end{figure}
%%%%%%%%%%%%%%%%%%%%%%%%%%%%%%%%%%%%%%%%%%%%%%%%%%%%%

A similar approach has been used e.g.,~in~\cite{Saharian:2004ih} for
infinitely thin Minkowski branes in a bulk AdS space; however, in our
case the contour we have to construct is complicated by the
presence of the poles which come from the bound state; and as we shall
see, it will be convenient to evaluate the bound state contribution
separately. Therefore, as it turns out, we shall only focus on the
total amplitude from now on.\footnote{The zeta function for the zero
mode is discussed in Appendix A for $d=2$ and $d=4$.}

In particular, we are primarily interested in calculating the mode
functions on the brane at $z=0\,(x=0)$, i.e.,
\begin{eqnarray}
\left.\frac{G(u,z)}{F(u)}\right|_{z=0}
%\nonumber\\
&=& \frac{
     \bigl( R^{iu\sigma}_{\nu}{}'(0) P^{iu\sigma}_{\nu}(0)
          - P^{iu\sigma}_{\nu}{}'(0) R^{iu\sigma}_{\nu}(0)
     \bigr)
     \bigl(R^{iu\sigma}_{\nu}{}'(x_L) P^{iu\sigma}_{\nu}(0)
          -P^{iu\sigma}_{\nu}{}'(x_L) R^{iu\sigma}_{\nu}(0)
     \bigr)
      }
      { -
    \bigl(
       R^{iu\sigma}_{\nu}{}'(0)   P^{iu\sigma}_{\nu}{}'(x_L)
  -    P^{iu\sigma}_{\nu}{}'(0)    R^{iu\sigma}_{\nu}{}'(x_L)
    \bigr)}
\nonumber \\
&=& 
   \frac{
     P^{iu\sigma}_{\nu}{}'(x_L) R^{iu\sigma}_{\nu}(0)
   -R^{iu\sigma}_{\nu}(x_L){}' P^{iu\sigma}_{\nu}(0)
      }
      { 
    \bigl(
    R^{iu\sigma}_{\nu}{}'(0)   P^{iu\sigma}_{\nu}{}'(x_L)
 -  P^{iu\sigma}_{\nu}{}'(0)   R^{iu\sigma}_{\nu}{}'(x_L)
   \bigr)}
\nonumber\\
&=&\frac{
     P^{iu\sigma}_{\nu}{}'(x_L) P^{-iu\sigma}_{\nu}(0)
   -P^{-iu\sigma}_{\nu}{}'(x_L) P^{iu\sigma}_{\nu}(0)
      }
      { 
    \bigl(
    P^{-iu\sigma}_{\nu}{}'(0)   P^{iu\sigma}_{\nu}{}'(x_L)
 -  P^{iu\sigma}_{\nu}{}'(0)   P^{-iu\sigma}_{\nu}{}'(x_L)
   \bigr)}\,,
%%%%%%%%%%%%%%%%%%%%%%%%%%%%%%%%%%%%%%%%%%%%
%=\frac{
%     \bigl(\alpha_{u}{} P^{iu\sigma}_{\nu}(0)
%          -\beta_{u} R^{iu\sigma}_{\nu}(0)
%     \bigr)
%     \bigl(\alpha_{u}{} P^{iu\sigma}_{\nu}(0)
%          -\beta_{u} R^{iu\sigma}_{\nu}(0)
%     \bigr)
%      }
%      { -\bigl(
%   \alpha_{u} P^{iu\sigma}_{\nu}{}'(x_L)
%  - \beta_{u} R^{iu\sigma}_{\nu}{}'(x_L)
%   \bigr)
% +\bigl(
%    \alpha_{u} P^{iu\sigma}_{\nu}{}'(0)
%  - \beta_{u} R^{iu\sigma}_{\nu}{}'(0)
%   \bigr)}
%\nonumber\\
%&=&\frac{
%     \bigl( R^{iu\sigma}_{\nu}{}'(0) P^{iu\sigma}_{\nu}(0)
%          - P^{iu\sigma}_{\nu}{}'(0) R^{iu\sigma}_{\nu}(0)
%     \bigr)
%     \bigl(R^{iu\sigma}_{\nu}{}'(x_L) P^{iu\sigma}_{\nu}(0)
%          -P^{iu\sigma}_{\nu}{}'(x_L) R^{iu\sigma}_{\nu}(0)
%     \bigr)
%      }
%      { -2
%    \bigl(
%       R^{iu\sigma}_{\nu}{}'(0)   P^{iu\sigma}_{\nu}{}'(x_L)
%  -    P^{iu\sigma}_{\nu}{}'(0)    R^{iu\sigma}_{\nu}{}'(x_L)
%    \bigr)}
%\nonumber \\
%&=& 
%   \frac{
%     P^{iu\sigma}_{\nu}{}'(x_L) R^{iu\sigma}_{\nu}(0)
%   -R^{iu\sigma}_{\nu}(x_L){}' P^{iu\sigma}_{\nu}(0)
%      }
%      { 2
%    \bigl(
%    R^{iu\sigma}_{\nu}{}'(0)   P^{iu\sigma}_{\nu}{}'(x_L)
% -  P^{iu\sigma}_{\nu}{}'(0)   R^{iu\sigma}_{\nu}{}'(x_L)
%   \bigr)}
%\nonumber\\
%&=&\frac{
%     P^{iu\sigma}_{\nu}{}'(x_L) P^{-iu\sigma}_{\nu}(0)
%   -P^{-iu\sigma}_{\nu}{}'(x_L) P^{iu\sigma}_{\nu}(0)
%      }
%      { 2
%    \bigl(
%    P^{-iu\sigma}_{\nu}{}'(0)   P^{iu\sigma}_{\nu}{}'(x_L)
% -  P^{iu\sigma}_{\nu}{}'(0)   P^{-iu\sigma}_{\nu}{}'(x_L)
%   \bigr)}\,,
\end{eqnarray}
where in the first step we used the Wronskian relation Eq. (\ref{Wronski}) and in the final step we specified the second mode function as
\begin{eqnarray}
R^{iq\sigma}_{\nu}(x)=-
\frac{\pi}{2i\sinh(\pi q\sigma)} P^{-iq\sigma}_{\nu}(x)\,.
\end{eqnarray} 
Two types of decomposition are possible:
\bea
&&\frac{
     P^{iu\sigma}_{\nu}{}'(x_L) P^{-iu\sigma}_{\nu}(0)
   -P^{-iu\sigma}_{\nu}{}'(x_L) P^{iu\sigma}_{\nu}(0)
      }
      { 
    P^{-iu\sigma}_{\nu}{}'(0)   P^{iu\sigma}_{\nu}{}'(x_L)
 -  P^{iu\sigma}_{\nu}{}'(0)   P^{-iu\sigma}_{\nu}{}'(x_L)}
\nonumber \\
&=&
\frac{P^{iu\sigma}_{\nu}(0)}{P^{iu\sigma}_{\nu}{}'(0)}
-\frac{P^{iu\sigma}_{\nu}{}'(x_L)}{P^{iu\sigma}_{\nu}{}'(0)}
\frac{2i\sinh(\pi u \sigma)}{\pi}
\frac{1}
{ P^{iu\sigma}_{\nu}{}'(0)   P^{-iu\sigma}_{\nu}{}'(x_L)
 -P^{iu\sigma}_{\nu}{}'(x_L) P^{-iu\sigma}_{\nu}{}'(0)}
\nonumber \\
&=&
\frac{P^{-iu\sigma}_{\nu}(0)}{P^{-iu\sigma}_{\nu}{}'(0)}
+\frac{P^{-iu\sigma}_{\nu}{}'(x_L)}{P^{-iu\sigma}_{\nu}{}'(0)}
\frac{2i\sinh(\pi u \sigma)}{\pi}
\frac{1}
{ P^{-iu\sigma}_{\nu}{}'(0)   P^{iu\sigma}_{\nu}{}'(x_L)
 -P^{-iu\sigma}_{\nu}{}'(x_L) P^{iu\sigma}_{\nu}{}'(0)}
\,.
\eea
It is important to note that the second term on the second line is
negligible in the $x_L\to 1$ limit on the upper half of the complex
$u$-plane, while the second term on the third line is negligible in
the same limit on the lower-half of
complex $u$-plane. Thus, in the single brane limit we use the first
term on the second and third lines as the single brane propagator on
the upper and lower half of the complex plane, respectively. 

In the single brane propagator given above,
$P^{iu\sigma}_{\nu}(0)/P^{iu\sigma}_{\nu}{}'(0)$ has poles that
are situated on the negative imaginary axis,~corresponding to purely
decaying modes, plus the bound state contribution at $u=i\nu/\sigma$.
However, as we mentioned above,
$P^{iu\sigma}_{\nu}(0)/P^{iu\sigma}_{\nu}{}'(0)$ is used for the upper
half of the complex plane and we need not worry about the purely decaying
modes. Thus, we only need to deal with the bound state mode at
$u=i\nu/\sigma$ in the calculation of the KK amplitude.     
Similarly, the exact opposite occurs 
for $P^{-iu\sigma}_{\nu}(0)/P^{-iu\sigma}_{\nu}{}'(0)$ 
and we only need to deal with the pole at $u=-i\nu/\sigma$.

%%%%%%%%%%%%%%%%% Discussions about contours %%%%%%%%%%%%%
The remaining problem then concerns the avoidance of the bound
state poles at  $u=\pm i{\nu}/{\sigma}$.
We avoid the bound state poles by deforming the contour to
$C'$, as depicted in Fig.~3, when we evaluate the KK amplitude. 
However, this contour gives a non-zero contribution (from the bound state 
poles) when taking the
Cauchy principal value on the imaginary axis.
This contribution simply corresponds to the subtraction of the bound
state from the total amplitude; we can calculate the bound
state amplitude separately, see Appendix A. 
Thus, it will be rather convenient for us to shift the contour over the 
upper pole to
$\tilde C$, as depicted in Fig.~4. This is equivalent to adding the bound 
state contribution with a counter-clockwise contour (the closed dotted line 
in Fig.~3) to $C'$.
Then, by integrating along the contour $\tilde C$ and subtracting the bound 
state contribution, we can obtain the desired KK amplitude.
This is the approach we shall take to evaluate the KK amplitude in
this article.

%%%%%%%%%%%%%%%%%%%%%%%%%%%%%%%%%%%%%%%%%%%%%%%%%%%%%%%%%%%%%%%%%%%%%%%%
\section{Kaluza-Klein Amplitude: $d=2$ Case}  

To demonstrate the method discussed in the previous section as simply as possible,~we first evaluate the amplitude of the quantum fluctuations on the brane  
for $d=2$. That is, we construct the zeta function for the
case of the two-sphere in the transverse dimensions
with one non-trivial bulk dimension.

\subsection{Amplitude of the KK modes}

The zeta function for total amplitude at the center of the wall is
\bea
\tilde \zeta(0,s)&=&
4\mu^{2(s-1)}
  \oint_{\tilde C} \frac{du}{2\pi i} 
  \frac{\sigma  u G(u,0)}{F(u)}
\sum_{j=0}^{\infty}\frac{(j+1/2)}
               {[u^2+(j+1/2)^2]^sH^{2s}}
\nonumber \\
&=&
 \frac{4\mu^{2(s-1)}}{\pi H^{2s}}
 \sin[\pi(s-1)] \,\,  
{\rm P}\int_{0}^{2\nu/\sigma}
   dU U  \frac{\sigma G(e^{\pi i/2}U,0)}{ F(e^{\pi i/2}U)}
  \sum_{j=0}^{\infty}
     \frac{(j+1/2)}
          {\bigl[U^2-(j+1/2)^2\bigr]^s}\,,
\label{twodsum}
\eea
where $U=e^{-\pi i/2}u$ and we use the property
\bea
\frac{ G(e^{\pi i/2}U, 0)}{ F(e^{\pi i/2}U)}
=\frac{ G(e^{-\pi i/2}U, 0)}{ F(e^{-\pi i/2}U)}
= \frac{P^{-U\sigma}_{\nu}(0)}
       {2P^{-U\sigma}_{\nu}{}'(0)}\,.\label{bulkpropagator}
\eea
Here, Roman ``${\rm P}$'' (not to be confused with the Legendre
function of the first kind)~means taking the Cauchy principal value in 
order to deal with
the pole at $U=\nu/\sigma$.  
In Fig.~4, the contribution from the anti-clockwise semi-circle around
$u=i\nu/\sigma$ cancels with that from the clockwise semi-circle around
$u=-i\nu/\sigma$. 

In the following, we shall divide the integral into two; i.e., 
for $U>2\nu/\sigma$ which we denote as the ``UV piece'' and that  
for $0<U<2\nu/\sigma$ which we denote as the ``IR piece.'' We
emphasize that the reason for this splitting is solely for technical
reasons and  that the choice of division has no physical
significance.
We can set the split at any value of $O(1)$.   

To begin with, for the UV piece we will use the following asymptotic
expansion formula, e.g., see Ref. \cite{Elizalde:1996zk}, for large
$U$, i.e.,
\begin{eqnarray}
\sum_{j=0}^{\infty} 
  \frac{(j+1/2)}
       {\bigl[U^2-(j+1/2)^2  \bigr]^s}
=- \frac{1}{2}U^{-2s+2}
\Bigl[
   \frac{1}{s-1}
  -\frac{1}{\Gamma(s)}
 \sum_{j=1}^{\infty}
   \frac{\Gamma(j+s-1)}
        {j!}
  U^{-2j}
   \partial_a  {\zeta}_{H}(-2j,a)\Big|_{a=1/2}
\Bigr]\,.
\label{summed}
\end{eqnarray}
Then, for the IR piece we employ the standard binomial expansions:
\begin{eqnarray}
\sum_{j=0}^{\infty}
     \frac{j+1/2}
          {\bigl[(j+1/2)^2-U^2\bigr]^s}
=\sum_{J=0}^{\infty}
    \frac{\Gamma(s+J)}{J!\Gamma(s)}
    U^{2J}
   \zeta_{H}(2s+2J-1,\frac{1}{2})\,,
\end{eqnarray} 
which is valid for the range $0<U<1/2$; 
while for the range $1/2<U<2\nu/\sigma < 1$ we must 
use \cite{Elizalde:1996zk, CREP} 
\begin{eqnarray}
\sum_{j=0}^{\infty}
     \frac{j+1/2}
          {\bigl[(j+1/2)^2-U^2\bigr]^s}
=\frac{1}{2}
\left(
  \frac{1}{(\frac{1}{4}-U^2)^s}  
+ \sum_{J=0}^{\infty}
    \frac{\Gamma(s+J)}{J!\Gamma(s)}
    U^{2J}
\left(
   2\zeta_{H}(2s+2J-1,\frac{1}{2})
  -\left(\frac{1}{2}\right)^{-2s-2J}
\right)
\right)\,.
\end{eqnarray}
Then, the total amplitude on the center at the wall is given by the
summation of both pieces
\bea
 \tilde \zeta (0,s)&=&
 \tilde \zeta_{\rm UV} (0,s)+\tilde \zeta_{\rm IR} (0,s)\,.
\label{totzeta}
\eea
%%%%%%%%%%%%
 
First, let us consider the analytic continuation of the UV piece
\bea
 \tilde \zeta_{\rm UV}(0,s)
&=&- \frac{2\mu^{2(s-1)}
 \sin[\pi (s-1) ]}
           {\pi H^{2s}}
\nonumber\\
&\times &
 \int^{\infty}_{2\nu/\sigma} dU
      \frac{\sigma  G(e^{\pi  i/2}U,0)}{F(e^{\pi i/2}U)} 
   U^{-2s+3} 
\left[
\frac{1}{s-1}
-\frac{1}{\Gamma(s)}
  \sum_{j=1}^{\infty}
 \frac{\Gamma(j+s-1)}{j!}U^{-2j}
  \partial_a \zeta_H (-2j,a)\Big|_{a=1/2}
\right]\,.
\nn    
\eea
Given the following relation \cite{ab}
\bea
\frac{P^{-U\sigma}_{\nu}(0)}
            {P^{-U\sigma}_{\nu}{}'(0)} 
=-{1\over 2}
\frac{\Gamma(-\frac{\nu}{2}+\frac{U\sigma}{2})
\Gamma( \frac{\nu}{2}+\frac{U\sigma}{2}+\frac{1}{2})}
{\Gamma( \frac{\nu}{2}+\frac{U\sigma}{2}+1)
\Gamma(-\frac{\nu}{2}+\frac{U\sigma}{2}+\frac{1}{2})}\,
\label{4gamma}
\eea
and by employing the asymptotic expansion for large $U$ of the Gamma
functions \cite{ab} we find the following asymptotic series,
which in $d$-dimensions is 
\bea
\frac{P^{-U\sigma}_{\nu}(0)}
            {P^{-U\sigma}_{\nu}{}'(0)} 
=\sum_{\ell=0}^{\infty}
  w_{\ell}(\sigma,\xi)U^{-1-2\ell}\,,   \label{legasymp}
\eea
where $\nu$ is given by Eq.~(\ref{nudef}) and
\bea
w_{0}(\sigma,\xi)
 & =&-\frac{1}{\sigma}\,,
\nonumber \\
w_1(\sigma,\xi)
&=&\frac{(2+\sigma(d-1))(1+d(-1+4\xi))}{8\sigma^2}
 =\frac{1}{2\sigma}\tilde V(0)
\,, 
\label{expanw}
\\
w_2(\sigma,\xi)
&=&-\frac{(2+\sigma(d-1))(1+d(-1+4\xi))\left(8 + 6\sigma\left( 1 + d\left( 
-1 + 4\xi \right)  \right)  + 
  3\sigma^2\left( -1 + d \right)\left( 1 + d\left( -1 + 4\xi \right)  
\right)\right)}
  {128\sigma^4}\nonumber\,.
\eea
The subtraction of the $w_0$ term just corresponds to that of
the trivial background, whereas the $w_1$ term
corresponds to the tadpole graph, see \cite{Olum:2002ra,
Graham:2002yr}.  
Here, we require only the term $w_0$, in order to regularize the
$d=2$ case.
For the $d=4$ case, terms up to $w_1$ are 
required.\footnote{In practice, for better numerical convergence we
subtract off more terms than are required to regularize the theory; thus,
we include $w_1$ for $d=2$ and $w_2$ for $d=4$, respectively.}

Thus, after analytic continuation to $s\to 1$, we obtain the UV amplitude
\bea
H^2\lim_{s\to 1}
\tilde{\zeta}_{\rm UV}(0,s)
&=&
-2
 \Bigl\{  
 \int^{\infty}_{2\nu/\sigma} dU U\,\sigma
\Bigl[\frac{P^{-U\sigma}_{\nu}(0)}
            {P^{-U\sigma}_{\nu}{}'(0)} 
           -\frac{w_0(\sigma,\xi)}{U}
   \Bigr]
   -w_0(\sigma,\xi)
          \left(\frac{2\nu}{\sigma}\right)
\Bigr\}\,.
\eea

As for the IR piece it is already finite in the limit $s\to 1$; however, 
because of the poles on the imaginary axis we make the principle value 
prescription, i.e.,
\bea
\tilde \zeta_{\rm IR}(0,s)
&=&\frac{4\mu^{2(s-1)}}{\pi  H^{2s}(-1)^s}
\sin[\pi(s-1)]\,
{\rm P} \int^{1/2}_0 dU
 \frac{\sigma U G(e^{\pi  i/2}U,0)}{F(e^{\pi i/2}U)} 
 \sum_{J=0}^{\infty}
    \frac{\Gamma(s+J)}{J!\Gamma(s)}
    U^{2J}
   \zeta_{H}(2s+2J-1,\frac{1}{2})
 \nonumber \\
& +&
 \frac{2\mu^{2(s-1)}}{\pi  H^{2s}(-1)^s}
\sin[\pi(s-1)]\,{\rm P}
 \int^{2\nu/\sigma}_{1/2}dU
  \frac{\sigma  G(e^{\pi  i/2}U,0)}{F(e^{\pi i/2}U)} 
\nonumber \\
&\times &
\left(
  \frac{1}{(\frac{1}{4}-U^2)^s}  
+ \sum_{J=0}^{\infty}
    \frac{\Gamma(s+J)}{J!\Gamma(s)}
    U^{2J}
\left(
   2\zeta_{H}(2s+2J-1,\frac{1}{2})
  -\left(\frac{1}{2}\right)^{-2s-2J}
\right)
\right)\,,
\label{2dir}
\eea
where if $2\nu/\sigma<1/2$ the second term is to be dropped.
Then, given the Laurent expansion of the Hurwitz zeta function
\begin{eqnarray}
\zeta_{H} (2s-1,\frac{1}{2})
=\frac{1}{2(s-1)}
 -\psi(1/2)+O(s-1)\,,\label{leurant2d}
\end{eqnarray}
we find that there is only a contribution from $J=0$ in both terms. Thus, 
in the limit $s\rightarrow 1$, we obtain 
\bea
H^2\lim_{s\to 1}\tilde\zeta_{\rm IR}  (0,s)
&=&
-{\rm P} \int^{2\nu/\sigma}_0  dU
      \frac{2\sigma  G(e^{\pi  i/2}U,0)}{F(e^{\pi i/2}U)} 
   U 
\nonumber\\
&=& 2\int^{2\nu/\sigma}_0  \frac{dU}{U-\nu/\sigma}
       \left(
 U \frac{\Gamma(-\nu/2+U\sigma/2+1)\Gamma(\nu/2+U\sigma/2+1/2)}
       {\Gamma(\nu/2+U\sigma/2+1)\Gamma(-\nu/2+U\sigma/2+1/2)}
   -\frac{1}{\sqrt{\pi}}     
       \frac{\nu}{\sigma}
    \frac{\Gamma(\nu+1/2)}{\Gamma(\nu+1)} 
     \right)
\,,
\nonumber
\\
\eea
where in the final step, we used the fact that
\bea
{\rm  P}\int_0^{2x_0}dx \frac{f(x)}{x-x_0}
&=& {\rm P}\int_{0}^{2x_0}dx\frac{f(x)}{x-x_0}
   -{\rm P}\int_{0}^{2x_0}dx\frac{f(x_0)}{x-x_0}
\nonumber \\
&=&\int^{2x_0}_0 dx\frac{f(x)}{x-x_0} 
  -\int^{2x_0}_0 dx\frac{f(x_0)}{x-x_0}\,,\label{Urawaza}
\eea
where $f(x)$ is an arbitrary regular function.
The second term, which is equal to zero, eliminates the singularity
at $x=x_0$ in the first term.
This technique will also be used for the $d=4$ case. 

Finally, we obtain the regularized total amplitude
\bea
\langle
\tilde \chi^2(0)
 \rangle_{\rm tot}
&=& \lim_{s\to 1}
\left( 
     \tilde \zeta_{\rm UV} (0,s)
    +\tilde \zeta_{\rm IR} (0,s)
    \right)\,.
\label{sumsum}
\eea
As discussed in the preceding section the KK amplitude is obtained by 
subtracting the bound state amplitude, which is evaluated in Appendix A,
\begin{eqnarray}
\langle \tilde \chi^2(0) \rangle_{\rm KK}
=  \langle \tilde \chi^2(0)\rangle_{\rm tot}
-\langle\tilde \chi^2(0)\rangle_{\rm bs}\,.
\label{KKamp2d}
\end{eqnarray}
Interestingly, the total amplitude $\langle \tilde \chi^2(0) \rangle_{\rm 
tot}$
does not depend on the renormalization scale $\mu$, whereas as shown
in Appendix $A$ the bound state contribution $\langle \tilde \chi^2(0) 
\rangle_{\rm bs}$ does depend on it.
Thus, the KK amplitude $\langle \tilde \chi^2(0) \rangle_{\rm KK}$
will also depend on $\mu$ as can be readily seen from Eq.~(\ref{KKamp2d}).
In other words, the dependence on $\mu$ in the bound state and KK
contribution cancels when they are summed up.

%%%%%%%%%%%%%%%%%%%%%%%%%%%%%%%%%%%%%%%%%%%%%%%%%%%%%%%%%%%%%%%%%%%%%%
\subsection{Results of numerical calculations}

The total amplitude $\langle \chi^2(0)\rangle_{\rm tot}$ is
shown in Fig.~5.  
For small thicknesses, the UV piece dominates the total amplitude. 
The leading order divergent behavior can be estimated as follows: by 
changing variables from $U$ to $x=U\sigma$, the UV piece can be written as 
\bea
 H^2\lim_{s\to 1} 
     \tilde \zeta_{\rm UV}(0,s)
=-2\left(
      \frac{1}{\sigma}\int^{\infty}_{2\nu} dx\,  x
       \left(
         \frac{P^{-x}_{\nu}(0)}{P^{-x}_{\nu}{}'(0)}
         +\frac{1}{x}
       \right)
     +\frac{2\nu}{\sigma}
  \right)\,.
\eea
As shown previously (see Fig. 2), $\nu/\sigma$ is almost independent of 
$\sigma$ and $\nu=O(\sigma)$ for $\sigma \ll 1$. Then by Taylor
expanding the Gamma functions in Eq. (\ref{4gamma}) about $\nu$ we
find that
\bea
 \frac{P^{-x}_{\nu}(0)}{P^{-x}_{\nu}{}'(0)}
=-\frac{1}{x}
 +\frac{\nu}{2x}
   \left(
      \psi(\frac{x}{2}) 
     -2\psi(\frac{x+1}{2})
     +\psi(\frac{x}{2}+1) 
   \right) 
 +O(\sigma^2)\,.
\eea 
Therefore, 
\bea
 H^2\lim_{s\to 1} 
     \tilde \zeta_{\rm UV}(0,s)
=-\frac{\nu}{\sigma}
     \int^{\infty}_{2\nu} dx
    \left(
    \psi(\frac{x}{2}) 
     -2\psi(\frac{x+1}{2})
     +\psi(\frac{x}{2}+1) 
    \right)  
   +O(\sigma^0).
\eea
In the case of $d=2$ the divergence arises only from the leading order.
Furthermore, for $x\gg 1$ the integrand behaves as $x^{-2}$
and thus, the contribution from the upper bound vanishes.
However, in the opposite limit, $x\ll 1$,
\bea
     \psi(\frac{x}{2})
     -2\psi(\frac{x+1}{2})
     +\psi(\frac{x}{2}+1) 
   =-\frac{2}{x}
    +\left(
       -2\gamma
       -2\psi(1/2) 
     \right)
   +O(x)\,,
\eea
where $\gamma=0.57721\cdots$ is Euler's constant, and therefore
\bea
 H^2\lim_{s\to 1} 
     \tilde \zeta_{\rm UV}(0,s)
=-\frac{2\nu}{\sigma}\ln\left(2\nu\right)
  +O(\sigma^0).
\eea
Thus, we find a positive logarithmic divergence in the thin wall limit.

The amplitude of the bound state is derived separately 
in Appendix A.
Here, we recapitulate the final result,
\bea
H^2
\langle
\tilde \chi^2(0)
 \rangle_{\rm bs}
&=& \frac{1}{\sigma}
 \Bigl( 2\ln\left(\frac{\mu}{H}\right)
       -2 \psi(1/2)
        -\delta_{\xi\,, 0} \left(\frac{1}{2}\right)^{-2}
 +    \sum_{J=1}^{\infty} 
 \Big[2\left(\frac{\nu}{\sigma}\right)^{2J}\zeta_H(2J+1,\frac{1}{2})
-\delta_{\xi\,, 0} \left(\frac{1}{2}\right)^{-2}\Big]
 \Bigr)
\nonumber \\
&\times&
\left(\int^{\infty}_0 dy \cosh^{-2\nu}(y) \right)^{-1}
\,.
\eea
In Fig.~6, the amplitude of the bound state is plotted as a
function of the brane thickness for each coupling, with $\mu=H$. 
Interestingly, the resulting amplitude is almost independent of the brane 
thickness $\sigma$ and still finite in the thin wall limit.

Thus, as expected, the divergence of the total amplitude in the thin wall 
limit
arises solely from the KK contribution.
Regardless, for finite values of $\sigma \sim 0.1$ the total amplitude 
settles down to finite positive values.
The result shows that the surface divergence for the KK modes can be 
regularized by introducing a finite brane thickness.
This is one of the main results of this article.

%%%%%%%%%%% Figure total amp  d=2 %%%%%%%%%%%%%%
\begin{figure}[htbp]
\begin{center}
  \begin{minipage}[t]{.45\textwidth} 
   \begin{center}
     \includegraphics[scale=.83]{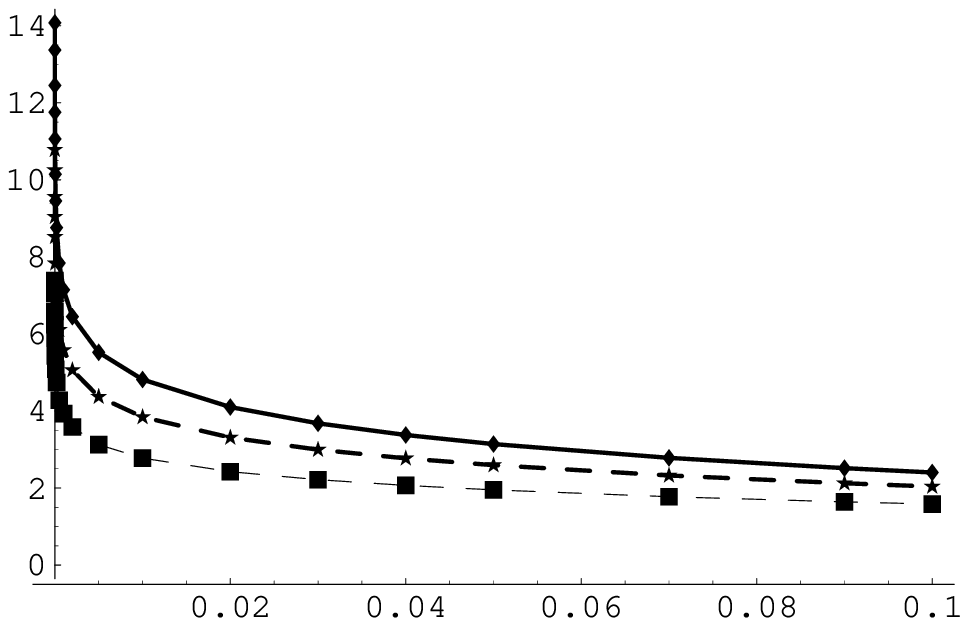}
\caption{
The total amplitude is shown as a
function of the brane thickness, $\sigma$, in the case of $d=2$.
The thick, thick-dashed and dashed curves
correspond to the cases of $\xi=0,1/32, 1/16$, respectively.}          
   \end{center}
  \end{minipage}
\hspace{0.5cm}
 \begin{minipage}[t]{.45\textwidth}
    \begin{center}
  \includegraphics[scale=.83]{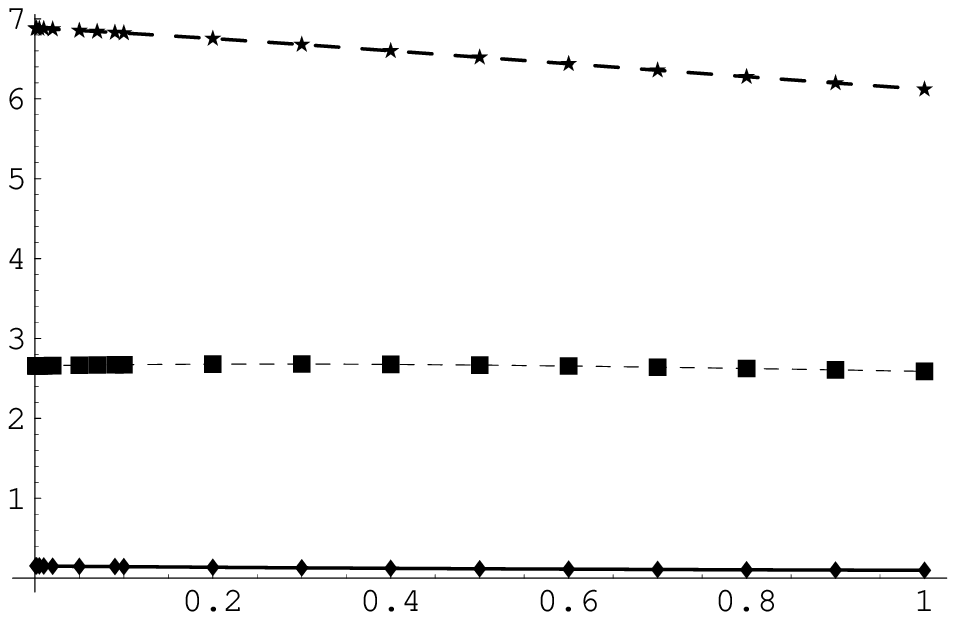} 
\caption{
The amplitude of the bound state is shown as a function of the
brane thickness, $\sigma$, in the case of $d=2$, with $\mu=H$.
The thick, thick-dashed  and dashed curves correspond to the cases 
of $\xi=0,1/32, 1/16$, respectively.}          
   \end{center}
  \end{minipage}
  \end{center}
\end{figure}

\begin{figure}[htbp]
\begin{center}
  \begin{minipage}[t]{.47\textwidth}
  \begin{center}
   \includegraphics[scale=.83]{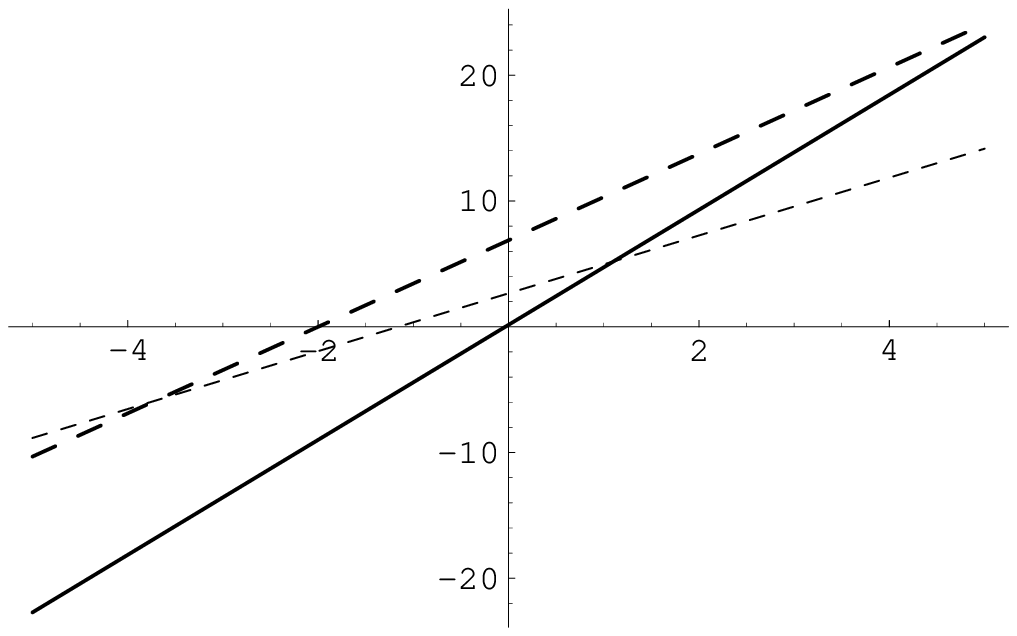}
\caption{
The running of the bound state is shown as a function of the
renormalization scale $\ln\,\mu$ in the case of $d=2$, with 
$\sigma=0.01$. The vertical and horizontal axes show the bound state 
amplitude and
$\log_{10}(\mu/H)$, respectively. 
The thick, thick-dashed  and dashed curves correspond
to the cases of $\xi=0,1/32, 1/16$, respectively.}          
 \end{center}
 \end{minipage}
\hspace{0.5cm}
 \begin{minipage}[t]{.47\textwidth}
  \begin{center}
 \includegraphics[scale=.83]{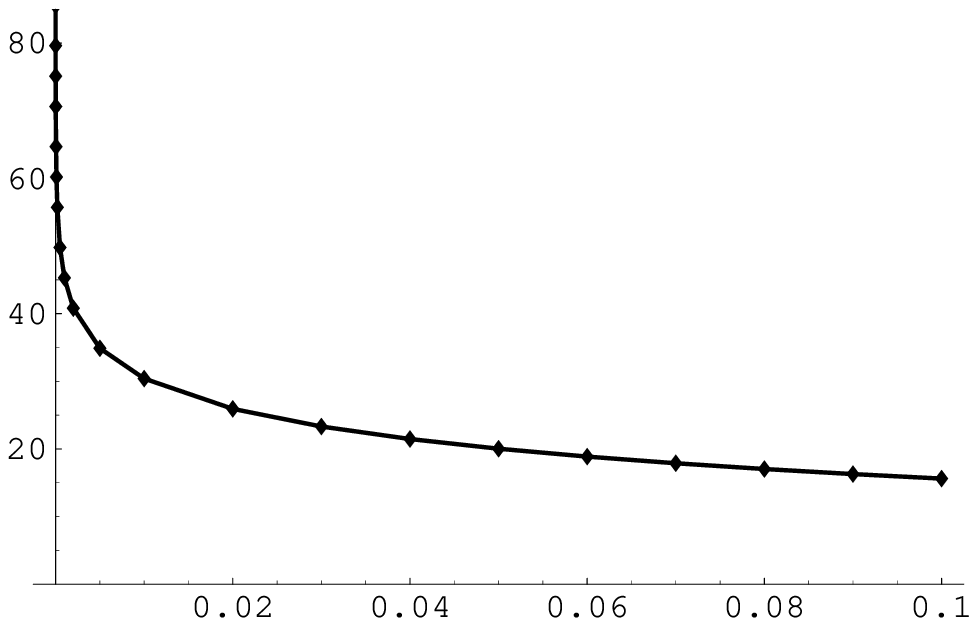}
\caption{
The relative amplitude of the KK modes to the bound state mode is
  shown as a function of the brane thickness, $\sigma$, for
  minimal coupling, $\xi=0$, for $d=2$, with $\mu=H$.} 
  \end{center}
 \end{minipage}
\end{center}
\end{figure}

The bound state amplitude depends on the choice of renormalization
scale, $\mu$.
In Fig.~7, the running of the scale is shown as a function
of $\mu$.
It is essentially proportional to $\ln\,\mu$.
The tilt becomes steeper for smaller coupling parameter $\xi$.  
There are several possible choices for the renormalization scale, for
example, one could choose the expansion rate of the brane $\mu=H$ or
another choice is the brane thickness $\mu=H/\sigma$. 
%However, a more realistic model might assume an explicit coupling of
%the $\chi$-field to other bulk or brane fields. In such a case we
%should then be able to impose an appropriate renormalization
%condition, but this is somewhat out of scope of this article. 
We still have no signature from braneworld today and therefore no quantity 
that we can renormalize into.
The renormalization scale $\mu$ should be determined by future observations
and/or experiments.
In this article we just plot the running of the scale and take the optimal
choice $\mu=H$ for cases where one has to make a choice.
Note that from Eq.~(\ref{KKamp2d}) the KK amplitude is also
proportional to $\ln(\mu)$ with negative tilts.

It is also interesting to compare the relative amplitude of the KK modes to 
the bound state mode. The relative amplitude is given by
\bea
r:=\frac{\langle  \chi^2(0) \rangle_{\rm KK}}
        {\langle  \chi^2(0) \rangle_{\rm bs}}\
  =\frac{\langle \tilde \chi^2(0) \rangle_{\rm  KK}}
        {\langle \tilde \chi^2(0) \rangle_{\rm bs}}
  =\frac{\langle \tilde \chi^2(0) \rangle_{\rm tot}}
        {\langle \tilde \chi^2(0) \rangle_{\rm bs} }
   -1\,,
\label{relative}
\eea
where in the final step we used Eq.~(\ref{KKamp2d}).
The result depends on the choice of the renormalization scale $\mu$
and brane thickness, $\sigma$. 
It is meaningful to show the plot for physically reasonable cases. 
As an example, in Fig.~8, we have plotted $r$ as a function of $\sigma$ for 
the minimally coupled case, $\xi=0$, i.e., for  tensor perturbations, with 
$\mu=H$.

%%%%%%%%%%%%%%%%%%%%%%%%%%%%%% Chapter 5 %%%%%%%%%%%%%%%%%%%%%
\section{Kaluza-Klein Amplitude: $d=4$ case}

In this section, we perform the calculation for the more realistic
case of $d=4$.
The calculation follows in an identical manner to the $d=2$ case, if
only for more tedium.  

\subsection{Amplitude of the KK modes}

In this case the degeneracy factor for the four-sphere ($d=4$) is
\bea
d_j = \frac{1}{3}\left(j+\frac{3}{2}\right)
                 \left(j+1\right)
              \left(j+2\right)\,
\eea
and hence, the zeta function for the total amplitude can be reduced to
\bea
\tilde \zeta(0,s)&=&
\frac{2}{3}\mu^{2(s-1)}
  \oint_{\tilde C} \frac{du}{2\pi i} 
  \frac{\sigma  u G(u,0)}{F(u)}
\sum_{j=0}^{\infty}\frac{(j+3/2)(j+1)(j+2)}
               {[u^2+(j+1/2)^2]^sH^{2s}}
\nonumber \\
&=&\frac{2\mu^{2(s-1)}}{3\pi H^{2s}} 
  \sin[\pi(s-1)]\,{\rm P}
\int^{\infty}_0
    dU \frac{\sigma  U G( e^{\pi i/2}U,0)}{F(e^{\pi i/2}U)}
   \sum_{j=0}^{\infty}
      \frac{(j+1)(j+2)(j+\frac{3}{2})}
            {\bigl[ U^2-(j+\frac{3}{2})^2\bigr]^s}\,,
\label{fourdsum}
\eea
where we used the properties of bulk propagator
Eq.~(\ref{bulkpropagator}). Again, Roman ``${\rm P}$'' represents taking 
the
Cauchy principal value to deal with the pole at $U=\nu/\sigma$.  
As for the $d=2$ case, we divide the total zeta function into a UV
piece, i.e., for $U>2\nu/\sigma$; and an IR piece, i.e.,
for $0<U<2\nu/\sigma$.  
Similarly, the choice of the division is just for later convenience.

To begin with, for the UV piece we shall use the asymptotic formula 
\cite{Elizalde:1996zk}
\bea
\frac{2}{3}\sum_{j=0}^{\infty}
   \frac{(j+3/2)(j+1)(j+2)}
        {\bigl[U^2-(j+3/2)^2\bigr]^s}
&=&\frac{2}{3}
  \sum_{j=0}^{\infty}
     \frac{(j+3/2)^3}{\bigl[U^2-(j+3/2)^2\bigr]^s}
-\frac{1}{6}
   \sum_{j=0}^{\infty}
     \frac{(j+3/2)}{\bigl[U^2-(j+3/2)^2\bigr]^s}\,
\nonumber \\
&=&(-1)^s
   \Bigg(-\frac{1}{12(s-1)(s-2)(s-3)}
    \partial_a^3\theta(-U^2,a,s-2)
+\frac{1}{12(s-1)}\partial_a  \theta(-U^2,a,s)
\nonumber \\
&-&\frac{1}{2(s-1)(s-2)}  \partial_a  \theta(-U^2,a,s-1)
\Bigg)_{a=3/2}\,
\nonumber \\
&=&-\frac{1}{12\Gamma(s)}
\Bigl[
  U^{-2(s-1)}
   \sum_{j=0}^{\infty}
    \frac{ \Gamma(j+s-1)}{j!}
    U^{-2j}
    \partial_a \zeta_{H}(-2j,a)  \Big|_{a=3/2}
\nonumber \\
&-&U^{-2(s-3)}
   \sum_{j=0}^{\infty}
    \frac{ \Gamma(j+s-3)}{j!}
    U^{-2j}
    \partial_a^3 \zeta_{H}(-2j,a)  \Big|_{a=3/2}
\nonumber \\
&+& 6   U^{-2(s-2)}
   \sum_{j=0}^{\infty}
    \frac{\Gamma(j+s-2)}{j!}
    U^{-2j}
    \partial_a \zeta_{H}(-2j,a)  \Big|_{a=3/2}
\Bigr]\,,
\eea
where
\bea
 \theta(q^2,a,s)
:=
 \sum_{j=0}^{\infty}
   \frac{1}{\bigl[(j+a)^2+q^2\bigr]^{s-1}}\,.
\eea

Then, for the IR piece, we use the following binomial expansions 
\cite{Elizalde:1996zk, CREP}:
\begin{eqnarray}
\frac{1}{3}
\sum_{j=0}^{\infty}
     \frac{(j+1)(j+2)(j+3/2)}
          {\bigl[(j+3/2)^2-U^2\bigr]^s}
=\frac{1}{3} \sum_{J=0}^{\infty}
    \frac{\Gamma(s+J)}{J!\Gamma(s)}
    U^{2J}
   \left(
   \zeta_{H}(2s+2J-3,\frac{3}{2})
 -\frac{1}{4}
    \zeta_{H}(2s+2J-1,\frac{3}{2})
   \right)\,,
\end{eqnarray}
which is valid for the range $0<U<3/2$; while for the range $3/2<U<5/2$ we 
must use
\begin{eqnarray}
&&\frac{1}{3} 
\sum_{j=0}^{\infty}
     \frac{(j+1)(j+2)(j+3/2)}
          {\bigl[(j+3/2)^2-U^2\bigr]^s} 
=
  \frac{1}{\left(\frac{9}{4}-U^2\right)^s}
 \nonumber\\
&&\qquad 
   +\sum_{J=0}^{\infty}
     \frac{\Gamma(s+J)}{J!\Gamma(s)} 
      U^{2J}
    \left[
      \frac{1}{3}
      \left(
         \zeta_{H}(2s+2J-3,\frac{3}{2})
      -\frac{1}{4}\zeta_{H}(2s+2J-1,\frac{3}{2})
      \right)
       -\left(\frac{3}{2}\right)^{-2s-2J}
    \right]\,,
\nonumber \\
\end{eqnarray}
and finally, for the range $5/2<U<\frac{2\nu}{\sigma}$ we have
\begin{eqnarray}
&&\frac{1}{3} 
\sum_{j=0}^{\infty}
     \frac{(j+1)(j+2)(j+3/2)}
          {\bigl[(j+3/2)^2-U^2\bigr]^s} 
=
  \frac{1}{\left(\frac{9}{4}-U^2\right)^s}
+\frac{5}{\left(\frac{25}{4}-U^2\right)^s}
 \nonumber\\
  &&\quad +
 \sum_{J=0}^{\infty}
     \frac{\Gamma(s+J)}{J!\Gamma(s)} 
      U^{2J}
    \Bigg[
      \frac{1}{3}
      \left(
         \zeta_{H}(2s+2J-3,\frac{3}{2})
      -\frac{1}{4}\zeta_{H}(2s+2J-1,\frac{3}{2})
      \right)
       -\left(\frac{3}{2}\right)^{-2s-2J}
%\nonumber \\
%    & -&
- 5\left(\frac{5}{2}\right)^{-2s-2J}
    \Bigg]\,.
\end{eqnarray}
The total zeta function is obtained from Eq. (\ref{totzeta}).

First, let us focus on the analytic continuation of the UV piece. 
Some simple manipulations lead to the following expression
\bea
\tilde \zeta_{\rm UV}(0,s)
&=&
\frac{1}{12}\frac{\mu^{2(s-1)}}{H^{2s}}
\int_{2\nu/\sigma}^{\infty} dU 
\frac{\sigma  G(e^{\pi i/2}U,0)}{F(e^{\pi i/2}U)}  
\nonumber\\
&\times&
\Bigl\{
U^{-2s+3}\sin[\pi(s-2)]
\Bigl[
   \frac{1}{s-1}\partial_a  \zeta_{H}(0,a)\Big|_{a=3/2}
   +\sum_{j=1}^{\infty}
    \frac{\Gamma(j+s-1)}{j!\Gamma(s)}U^{-2j}
    \partial_a \zeta_{H}(-2j,a)  \Big|_{a=3/2}
   \Bigr]
\nonumber \\
&-&
U^{-2s+7}\sin[\pi(s-4)]
\Bigl[
\frac{1}{(s-1)(s-2)(s-3)}\partial_a^3  \zeta_{H}(0,a)  \Big|_{a=3/2}
+\frac{U^{-2}}{(s-1)(s-2)}\partial_a^3\zeta_{H}(-2,a)  \Big|_{a=3/2}
\nonumber  \\
&&
+\frac{U^{-4}}{2(s-1)}\partial_a^3\zeta_{H}(-4,a)  \Big|_{a=3/2}
+\sum_{j=3}^{\infty}
    \frac{\Gamma(j+s-3)}{j!\Gamma(s)}U^{-2j}
    \partial_a^3 \zeta_{H}(-2j,a)  \Big|_{a=3/2}
\Bigr]
\nonumber \\
&-&
6U^{-2s+5}\sin[\pi(s-3)]
\Bigl[
 \frac{1}{(s-1)(s-2)}\partial_a \zeta_{H}(0,a)  \Big|_{a=3/2}
+\frac{U^{-2}}{s-1}\partial_a\zeta_{H}(-2,a)  \Big|_{a=3/2}
\nonumber \\
&+&\sum_{j=2}^{\infty}
    \frac{\Gamma(j+s-2)}{j!\Gamma(s)}U^{-2j}
    \partial_a \zeta_{H}(-2j,a)  \Big|_{a=3/2}
\Bigr]
\Bigr\}\,.
\eea
Like for the $d=2$ case, after analytic continuation to $s\to
1$, this leads to
\bea
H^2\lim_{s\to 1}  \tilde \zeta_{\rm UV}(0,s)
&=&
-\frac{1}{3}\sigma 
\left(
\int_{2\nu/\sigma}^{\infty}
dU U^3
 \Bigl( \frac{P^{-U\sigma}_{\nu}(0)}{P^{-U\sigma}_{\nu}{}'(0)}-
\sum_{\ell=0}^{1}w_{\ell}(\sigma,\xi)
     U^{-1-2\ell}\Bigr)
+
  \sum_{\ell=0}^{1}\frac{2^{3-2\ell}
                    w_{\ell}(\sigma,\xi)}{2\ell-3}
  \left(\frac{\nu}{\sigma}\right)^{3-2\ell}
\right)
\nonumber \\
&+&
 \frac{1}{12}
\sigma
\left(
\int_{2\nu/\sigma}^{\infty}
dU U
 \Bigl(  \frac{P^{-U\sigma}_{\nu}(0)}{P^{-U\sigma}_{\nu}{}'(0)}
     -w_{0}(\sigma,\xi) U^{-1}\Bigr)
-2 w_{0}(\sigma,\xi)
     \left(\frac{\nu}{\sigma}\right)
\label{UV4d}
\right)\,,
\eea
where $w_{\ell}(\xi,\sigma)$ are the coefficients of the asymptotic 
expansion
in Eq.~(\ref{legasymp}) given by  Eq.~(\ref{expanw}), for $d=4$.

The IR piece is already finite for $s\to 1$, and some calculation 
shows that
\bea
\tilde  \zeta_{\rm IR}(0,s)
&=&
 \frac{2\mu^{2(s-1)}}{3\pi H^{2s}(-1)^s}\sin[\pi(s-1)]
  \, {\rm P}
    \int^{3/2}_0 dU
    \frac{\sigma UG(e^{\pi i/2}U,0)}{F(e^{\pi i/2}U)}
\nonumber \\
&\times&
   \sum_{J=0}^{\infty}
    \frac{\Gamma(s+J)}{J!\Gamma(s)}    U^{2J}
   \left(
   \zeta_{H}(2s+2J-3,\frac{3}{2})
 -\frac{1}{4}
    \zeta_{H}(2s+2J-1,\frac{3}{2})
   \right)
\nonumber\\
&+& 
  \frac{2\mu^{2(s-1)}}{\pi H^{2s}(-1)^s}\sin[\pi(s-1)]
   \, {\rm P}
    \int^{5/2}_{3/2} dU
    \frac{\sigma U G(e^{\pi i/2}U,0)}{F(e^{\pi i/2}U)}
\Biggl(
 \frac{1}{\left(\frac{9}{4}-U^2\right)^s}
 \nonumber\\
  & +&\sum_{J=0}^{\infty}
     \frac{\Gamma(s+J)}{J!\Gamma(s)} 
      U^{2J}
    \left[
      \frac{1}{3}
      \left(
         \zeta_{H}(2s+2J-3,\frac{3}{2})
      -\frac{1}{4}\zeta_{H}(2s+2J-1,\frac{3}{2})
      \right)
       -\left(\frac{3}{2}\right)^{-2s-2J}
    \right]
\Biggr)
\nonumber \\
&+& \frac{2\mu^{2(s-1)}}{\pi H^{2s}(-1)^s}\sin[\pi(s-1)]
   \, {\rm P}
    \int^{2\nu/\sigma}_{5/2} dU
    \frac{\sigma U G(e^{\pi i/2}U,0)}{F(e^{\pi i/2}U)}
\Bigg(
  \frac{1}{\left(\frac{9}{4}-U^2\right)^s}
+\frac{5}{\left(\frac{25}{4}-U^2\right)^s}
 \nonumber\\
  & +&\sum_{J=0}^{\infty}
     \frac{\Gamma(s+J)}{J!\Gamma(s)} 
      U^{2J}
    \Bigg[
      \frac{1}{3}
      \left(
         \zeta_{H}(2s+2J-3,\frac{3}{2})
      -\frac{1}{4}\zeta_{H}(2s+2J-1,\frac{3}{2})
      \right)
       -\left(\frac{3}{2}\right)^{-2s-2J}
     - \frac{5}{\left(5/2\right)^{2s+2J}}
    \Bigg]
  \Biggr)\,.
\nonumber \\
&&
\eea 
Note that the number of terms depends on the range of $U$. For $3/2<
2\nu/\sigma<5/2$ the third term should be dropped; while both the
second and third terms should be dropped if $2\nu/\sigma<3/2$.

In the $s\to 1$ limit, as before, just terms with leading order
\begin{eqnarray}
\zeta_{H}(2s-1,\frac{3}{2})
              =\frac{1}{2(s-1)}
              -\psi(3/2)
              +O(s-1)\,,\label{leurant4d}
\end{eqnarray}
contribute to the resulting IR amplitude.
Thus, taking the limit $s\to 1$, we obtain the IR amplitude as
\bea
H^2
\lim_{s\to 1} \tilde \zeta_{\rm IR} (0,s)
&=&
-\frac{1}{3}\sigma\, {\rm P}\int^{2\nu/\sigma}_0 dU 
\frac{U^3G(e^{\pi i/2}U,0)}{F(e^{\pi i/2}U)}
+\frac{1}{12}\sigma \,{\rm P}\int^{2\nu/\sigma}_0 dU 
\frac{UG(e^{\pi i/2}U,0)}{F(e^{\pi i/2}U)}
\nonumber \\
&=&
\frac{1}{3}\int^{2\nu/\sigma}_0 dU 
\frac{1}{U-\frac{\nu}{\sigma}}
\left(
U^3\frac{
\Gamma(-\frac{\nu}{2}+\frac{U\sigma}{2}+1)
\Gamma(\frac{\nu}{2}+\frac{U\sigma}{2}+\frac{1}{2})}
{\Gamma(\frac{\nu}{2}+\frac{U\sigma}{2}+1)
\Gamma(-\frac{\nu}{2}+\frac{U\sigma}{2}+\frac{1}{2})}
-\frac{1}{\sqrt{\pi}}
 \left(\frac{\nu}{\sigma}\right)^3 
 \frac{\Gamma(\nu+\frac{1}{2})}{\Gamma(\nu+1)}
\right)
\nonumber \\
&-&
\frac{1}{12}\int^{2\nu/\sigma}_0 dU 
\frac{1}{U-\frac{\nu}{\sigma}}
\left(
U\frac{
\Gamma(-\frac{\nu}{2}+\frac{U\sigma}{2}+1)
\Gamma(\frac{\nu}{2}+\frac{U\sigma}{2}+\frac{1}{2})}
{\Gamma(\frac{\nu}{2}+\frac{U\sigma}{2}+1)
\Gamma(-\frac{\nu}{2}+\frac{U\sigma}{2}+\frac{1}{2})}
-\frac{1}{\sqrt{\pi}}
 \left(\frac{\nu}{\sigma}\right)
 \frac{\Gamma(\nu+\frac{1}{2})}{\Gamma(\nu+1)}
\right)
\,,
\eea
where in the final step Eq. (\ref{Urawaza}) was used.

Finally, we obtain the total regularized amplitude from Eq.~(\ref{sumsum}).
Furthermore, the KK amplitude is obtained by subtracting the bound state
amplitude (evaluated in Appendix A) obtained from Eq. (\ref{KKamp2d}).
Note that the KK amplitude again has a dependence on the renormalization 
scale $\mu$.

%%%%%%%%%%%%%%%%%%%%%%%%%%%%%%%%%%%%%%%%%%%%%%%%%%%%%%%%%%%%
\subsection{Results of numerical calculations}

In Fig.~9, a numerical plot of $\langle \chi^2(0)\rangle_{\rm tot}$ 
is shown. Again, the divergence for the thin wall limit can be seen.
The power of the divergence can be estimated as follows: the dominant 
contribution in the thin wall limit comes from the first
term on the right hand side of Eq. (\ref{UV4d}). 
By changing variables to $x=U\sigma$ and following the same steps as 
for $d=2$, we obtain
\bea
&&\sigma 
\left(
\int_{2\nu/\sigma}^{\infty}
dU U^3
 \Bigl( \frac{P^{-U\sigma}_{\nu}(0)}{P^{-U\sigma}_{\nu}{}'(0)}-
\sum_{\ell=0}^{1}w_{\ell}(\sigma,\xi)
     U^{-1-2\ell}\Bigr)
+ \sum_{\ell=0}^{1}\frac{2^{3-2\ell}
                    w_{\ell}(\sigma,\xi)}{2\ell-3}
  \left(\frac{\nu}{\sigma}\right)^{3-2\ell}
\right)
\nonumber \\
&=&
  \frac{1}{\sigma^2}
  \int^{\infty}_{2\nu}
   dx\,
  \left(
     \frac{1}{2}x^2
     \left(\frac{\nu}{\sigma}\right)
     \left(
      \psi(\frac{x}{2})
     -2\psi(\frac{x+1}{2})
     +\psi(\frac{x}{2}+1) 
     \right)
     -\frac{(-3+16\xi)}{4}    
  \right)
+O(\sigma^{-1})\,.
\eea
In this case, the contribution from the lower bound of the
integration does not contribute to any power of $\sigma$.
Thus, in the thin wall limit the regularized amplitude diverges as
$\sigma^{-2}$. This is more divergent than the case of $d=2$ and is related 
to the fact that in higher dimensions we need higher powers of UV 
subtraction.

The amplitude of the bound state is calculated in Appendix A 
and is
\bea
H^2\langle  \tilde \chi^2(0)\rangle_{\rm bs}
&=&
\frac{1}{2\sigma}
           \left(
                \int^{\infty}_0 dy \cosh^{-2\nu}(y)
           \right)^{-1}
\nonumber  \\
&\times&
\Bigl\{
  \left(
            -\frac{1}{6}
           +\frac{2}{3}
          \left(\frac{\nu}{\sigma}\right)^2
    \right)
    \ln\left(\frac{\mu}{H}\right)     
+ \frac{2}{3}\zeta_{H}(-1,\frac{3}{2})
     +\frac{1}{6}\psi(3/2)
-\left(\frac{\nu}{\sigma}\right)^2
\left(
    -\frac{1}{3}
   +\frac{2}{3}\psi(3/2)
   +\frac{1}{6}\zeta_H (3,\frac{3}{2})
\right)
\nonumber \\  
 &+&\frac{2}{3}
      \sum_{J=2}^{\infty}
    \left(\frac{\nu}{\sigma}\right)^{2J}
 \left(  
 \zeta_H(2J-1,\frac{3}{2}) 
 -\frac{1}{4}  \zeta_H(2J+1,\frac{3}{2}) 
 \right)
     -2\sum_{J=0}^{\infty}\delta_{\xi \,,0}
       \left(\frac{3}{2}\right)^{-2} 
\Bigr\}\,.
\label{bsamp4d}
\eea
This is plotted in Fig.~10 and we see that the bound state is almost 
independent of the brane thickness 
and still finite in the thin wall limit.
Thus, like for $d=2$, the divergence in the total amplitude arises solely 
from the KK modes.

%%%%%%%%%%%%%%%  Amp of bs d=4  %%%%%%%%%%%%%%%%%%%%%%
\begin{figure}[tbp]
\begin{center}
 \begin{minipage}[b]{.47\textwidth}
  \begin{center}
   \includegraphics[scale=.83]{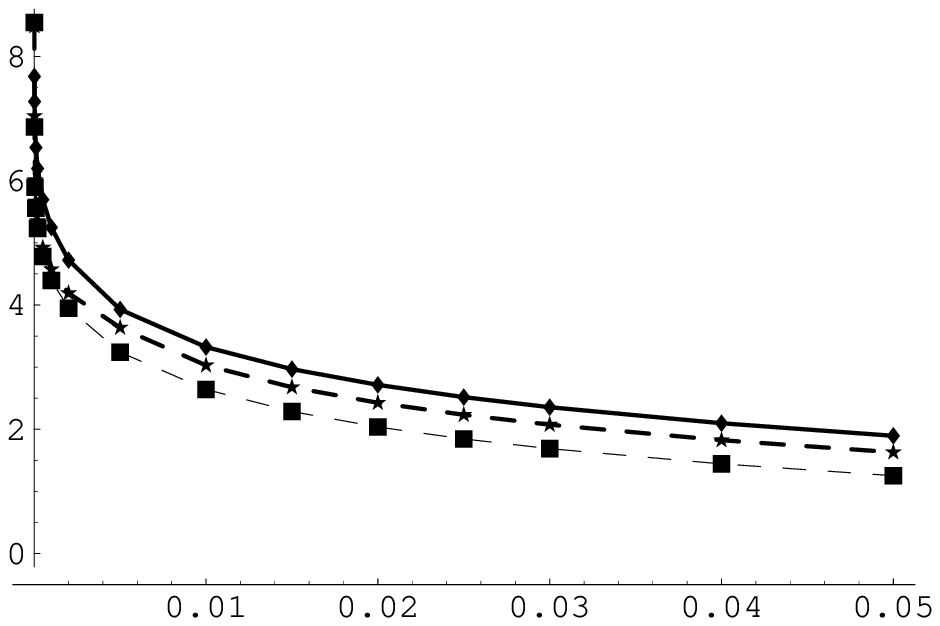}
\caption{ 
The total amplitude is shown as a
function of the brane thickness, $\sigma$, in the case of $d=4$.
The vertical axis shows $\log_{10}(-H^2\langle
\chi^2(0)\rangle_{\rm KK})$.
The thick, thick-dashed and dashed curves
correspond to the cases of $\xi=0,3/32, 3/20$, respectively.}          
    \end{center}
   \end{minipage}
\hspace{0.5cm}
   \begin{minipage}[b]{.47\textwidth}    
    \begin{center}
  \includegraphics[scale=.83]{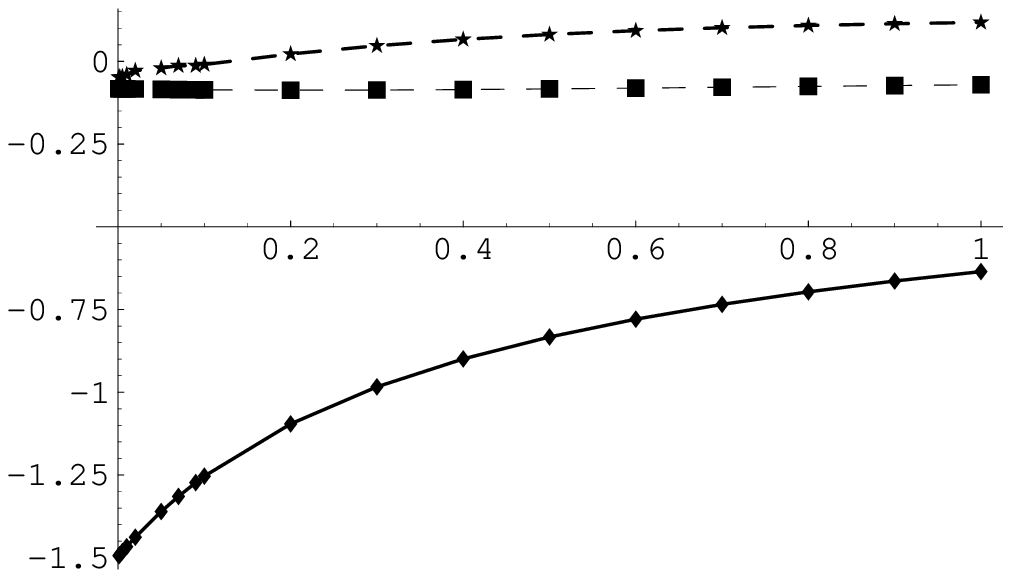}
\caption{ 
The amplitude of the bound state mode is shown as a
function of the brane thickness, $\sigma$, in the case of $d=4$, with
$\mu= H$.
The thick, thick-dashed and dashed curves correspond to the cases of
$\xi=0, 3/64, 3/32$, respectively.}
   \end{center}
   \end{minipage}
   \end{center}
\end{figure}
%%%%%%%%%%%%%%%%%%%%%%%%%%%%%%%%%%%%%%%%%%%%%%%%%%%%

Again, the amplitude depends on the choice of the renormalization scale 
$\mu$.
%%%%%%%%%%%%%%%%%%%new comments%%%%%%%%%%%%%%%%%
As we stated in the previous section, we have no 
signature from braneworld as yet and no way to determine the renormalization scale.
%%%%%%%%%%%%%%%%%%%%%%%%%%%%%%%%%%%%%%%%%%%%%%%%%
In certain cases the amplitude of the bound state can become negative.
In Fig.~11, the running of the bound state amplitude is shown 
as a function of $\mu$.
It is basically the same as the case of $d=2$; however, a new feature
is that negative tilts of the running are realized for larger values
of coupling $\xi$ which satisfy
\bea
\xi> \frac{2\sigma+1}{4(3\sigma+2)}\,,\label{negtilt}
\eea
as can be seen from Eq.~(\ref{bsamp4d}).
The critical coupling parameter in Eq.~(\ref{negtilt}) is smaller 
than conformal coupling, $\xi_c=3/16$, for any choice of brane thickness, 
$\sigma$. This fact means that there
always exist coupling parameters which realize negative tilts of the 
running.

%%%%%%%%%%% Figure  running of bs amp in d=4   %%%%%%%%%%%%%%
 
\begin{figure}[htbt]
\begin{center}
  \begin{minipage}[t]{.47\textwidth}
 \begin{center}
  \includegraphics[scale=.83]{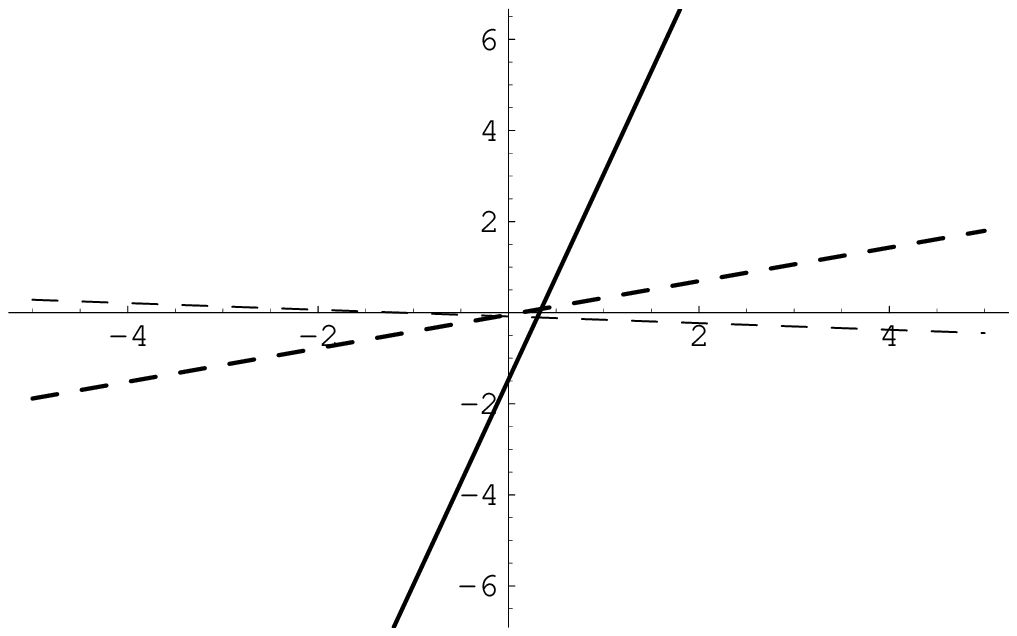}
\caption{
The running of the amplitude of the bound state is shown as a function of 
the brane thickness, $\sigma$, in the case of $d=4$, with $\sigma=0.01$.
The vertical and horizontal axes show the bound state amplitude and
$\log_{10}(\mu/H)$, respectively. 
The thick, thick-dashed  and dashed curves correspond to the cases of 
$\xi=0,
3/32, 3/20$, respectively.}          
  \end{center}
\end{minipage}
\hspace{0.5cm}
\begin{minipage}[t]{.47\textwidth}
\begin{center}
  \includegraphics[scale=.83]{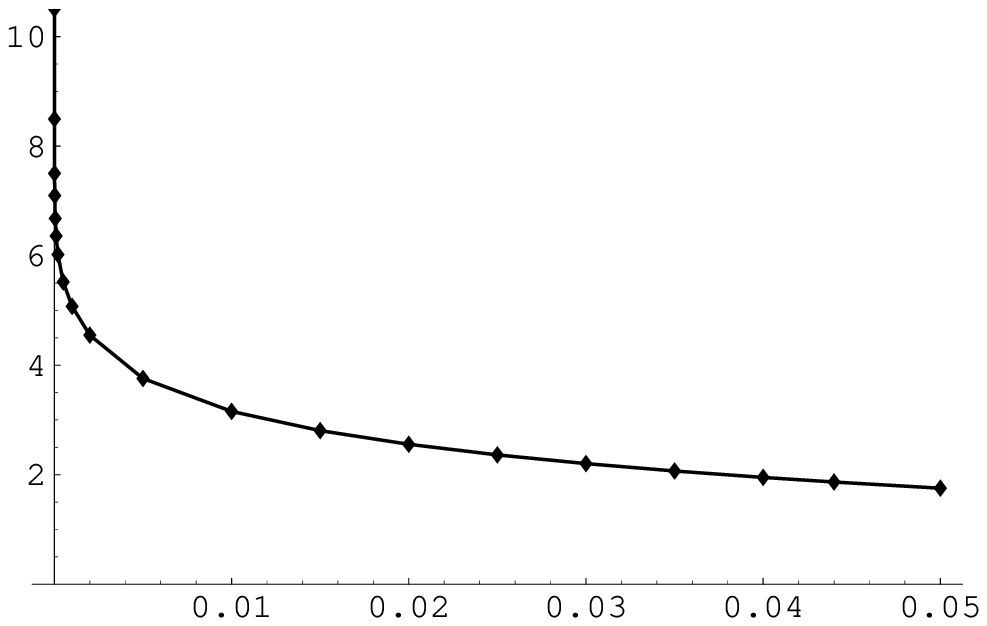}
\caption{
The relative amplitude of the KK modes to the bound state mode is shown as 
a function of the brane thickness, $\sigma$, for the minimal 
coupling, $\xi=0$, for $d=4$, with $\mu= H$.
The vertical axis shows $\log_{10}|r(\mu)|$ given by
Eq.~(\ref{relative}).}
\end{center}
 \end{minipage}
 \end{center}
\end{figure}
%%%%%%%%%%%%%%%%%%%%%%%%%%%%%%%%%%%%%%%%%%%%%%%%%%%%%%%%%%

The relative amplitude of the KK to bound state ratio,
defined by Eq. (\ref{relative}), depends on the choice of
renormalization scale $\mu$.
Again as one of the possible physical choices, in Fig.~12, we plot the 
relative amplitude in the case of $\mu=H$ for minimal
coupling, $\xi=0$.

%%%%%%%%%%%%%%%%%%%%%%%%%%%%%%%%%%%%%%%%%%%%%%%%%%%%%%%%%%
\section{Summary and discussion}

We discussed the quantum fluctuations in a thick brane model 
in order to examine whether or not a finite brane thickness can act as a 
{\it natural} cut-off of for the Kaluza-Klein (KK) mode spectrum. The thick brane model we examined was supported by a scalar field
with an axion-type potential. The thin brane limit of this model is
smoothly matched to the system of a de Sitter (dS) brane in a Minkowski 
bulk. 
As we showed for general $d+1$ dimensions, this model is classically
stable both against classical tensor and
scalar metric perturbations (see Appendix B).

Next, we introduced a test quantized scalar field, $\chi$, into this
model and calculated its amplitude.
This scalar field is assumed to have a zero bulk mass and non-zero
coupling to the background domain wall geometry.
For simplicity we ignore any explicit coupling of $\chi$ to the
geometry; namely, to the geometry or the supporting scalar field $\phi$. 
A particularly interesting case is that for minimal coupling, $\xi=0$,
which (for $d=4$) is equivalent to that of the tensor perturbations
for a possible low energy realization of general relativity. 

The squared amplitude of the KK modes was evaluated by using the 
dimensional
reduction approach developed in Ref.~\cite{Naylor:2004ua}. 
In order to obtain a well-posed quantum field theory, we introduced
a regulator boundary into the set up implying that the KK modes become discrete.
We worked in Euclidean space rather than the original Lorentzian space. 
For the purpose of the calculation of the amplitude, we used the
{\it local} zeta function method, where by ``{\it local}''
we mean that the quantity is integrated out over the volume of the 
$d$-sphere, $S^d$
(which is $d$-dimensional dS spacetime in Euclidean space), but not
over the extra-dimension, $z$. 
This quantity is described by a summation over all the KK modes and
internal modes associated with dS space.
Then, given the residue theorem, we can convert the summation of
all the KK modes into a contour integral representation.
The KK modes correspond to poles on the real axis and the contour $C$ is
taken to enclose them, which is depicted in Fig.~3.
Furthermore, as in Fig.~3, it can be deformed to the contour $C'$.
Then, by keeping the index of the zeta function $s$ large, we can obtain
an integral along the imaginary axis with two clockwise  
semi-circles to avoid the bound state poles.
Such a contour, $C'$, corresponds to the subtraction of the bound state 
contribution from the total amplitude, i.e., the KK contribution.
However, for technical reasons it was convenient to integrate along the 
contour $\tilde C$, depicted in Fig.~4, by adding the bound state
contour onto the contour $C'$.
The quantity obtained from the integration along $\tilde C$ is
the total amplitude.
Thus, after subtracting the bound state part, evaluated in Appendix A,
we were able to get the desired KK amplitude.

The bulk propagator can then be decomposed into two parts, where one
is regulator-brane independent and the other is regulator dependent.
Then, by sending the regulator-brane away from the domain wall to
infinity, the regulator dependent part vanishes and we can take a
well-defined single brane limit.

We decomposed the total zeta function into a high
frequency (UV) piece and a low frequency (IR) piece to accomplish a
successful regularization. Then, we summed these pieces up to get the total 
amplitude.

As an exercise, we first calculated the quantum fluctuations for the
$d=2$ case. 
For extremely small thicknesses the UV piece dominates and exhibits
a logarithmic divergence in the thin wall limit. For larger
thicknesses the total amplitude settles down to finite positive values.
Furthermore, we also calculated the amplitude of the bound state, which 
depends on
the renormalization scale. However, for a fixed scale
we find that its amplitude is almost independent of the brane
thickness and is finite in the thin wall limit.
Thus, only the KK modes lead to surface divergences in this limit.
In other words, the KK amplitude is regularized by the presence of a
finite brane thickness.
This is one of the main results in this article.
We also discussed the running of the bound state amplitude and 
(of particular physical significance) the relative amplitude of the
KK modes to the bound state contribution.

Then, we calculated the same quantities in the more realistic case 
of $d=4$. The main results are very similar to the $d=2$ case and need
not be recapitulated. The main difference is the divergent behavior in
the thin wall limit. We showed that the KK amplitude
diverges quadratically, as opposed to logarithmically for $d=2$. We also 
showed
that the KK amplitude has an overall negative magnitude.\footnote{This
was for our particular choice of renormalization scale, $\mu=H$;
however, in general it has an overall negative magnitude unless one
chooses $\mu$ to be very large.}
Regardless, for $d=4$ the amplitude of the KK modes is also free of
surface divergences for a finite brane thickness.
%%%%%%%%%%%%%%%new comment%%%%%%%%%%%%%%%%%%%
We have no signature from braneworld today
and the renormalization scale should be determined by future observations
and/or experiments.
%%%%%%%%%%%%%%%%%%%%%%%%%%%%%%%%%%

%%%%%%%%%%%%new paragraph(another thick brane model)%%%%%%%%%%%%%%%%%%%%%%
In this article we investigated the quantum fluctuations in a 
particular model of thick braneworld. However, the qualitative behaviour
of the quantum fluctuations should be independent of the choice of the model.
This can be understood as follows.
For a thick brane model, for which the spacetime is smooth everywhere, 
there will be no divergence. 
Now if we look at the behaviour of the background solution,
Eq. (\ref{thickbranesolution}), 
when $\sigma$ is sufficiently small, 
we find $\phi\sim \phi_0 (Hz/\sigma)$ for $Hz\ll \sigma$.
 This is a very general behaviour that one finds at the center of any
 domain wall solution, independent of the global features of the bulk potential.
 Thus the divergence in the thin wall limit is due to the
spacetime singularity caused by the divergence of $d\phi/dz=0$ at $z=0$,
which is common to any thick brane model supported by a bulk scalar field.
%%%%%%%%%%%%%%%%%%%%%%%%%%%%%%%%%%

%%%%%%%%%%%%%new paragraph (tensor perturbations)%%%%%%%%%%%%%%%%%
We are also interested in the amplitude of the bulk tensor metric perturbations, 
i.e., gravitons. As we mentioned in Sec.~III A
 the wave equation for the massless, minimally coupled,
 test scalar field discussed in this article, is equivalent to that of 
the tensor perturbations and we therefore expect a similar result;
 though an explicit demonstration is left for future work. 
%%%%%%%%%%%%%%%%%%%%%%%%%%%%%%%%%%%%%%%%%%%%%%%%%%%%%%%%%%%%%%%

In summary, in this article we have shown that a finite brane
thickness acts as a natural cut-off for the KK spectrum.
This fact implies that brane models which have a finite thickness
are more plausible than infinitesimally thin ones.

Some issues remain. In this article, we focused on the fluctuation
amplitude just at the center of the wall $z=0$, i.e.,~the brane position in
the thin wall limit, but the configuration of the fluctuations in the
bulk, i.e.,~the
$z$ dependence of the amplitude, especially close to the brane, is also
important. 
Moreover, we should also evaluate the Hamiltonian density,
which is an important quantity in its own right, particularly
concerning the back-reaction of the KK modes  (especially for small
brane thicknesses).
Employing the method developed here, it is possible to evaluate such
quantities \cite{Minamitsuji:2005sm}.

The more realistic case of a thick brane model which is embedded in an
asymptotically AdS spacetime\footnote{In the high
energy limit, $H\ell\gg 1$ ($\ell$ is the AdS curvature radius), the
bulk effectively becomes Minkowski and hence should reduce to the model
presented here.} or into a bulk with higher dimensions,\footnote{For example, the 
quantum effects of thin branes with higher spatial
dimensions and alike were recently discussed in 
\cite{Scardicchio:2005hh,Saharian:2005xf,Saharian:2005vm}.} 
is also of interest.
We hope to report on such topics in the near future.

%%%%%%%%%%%%%%%%%%%%%%%%%%%%%%%%%%%%%%%%%%%%%%%%%%%%%%%%
\begin{acknowledgments}

This work is supported in part by Monbukagakusho Grant-in-Aid for Scientific 
Research(S) No. 14102004 and (B) No.~17340075. 
\end{acknowledgments}

%%%%%%%%%%%%%%%%%%%%%%%%%%%%%%%%%%%%%
%%%%%%%%%%%%%%%%%%%%%%%%%%%%%%%%%%%%%%%%%%%%%%%%%%%%%%%%%%%%%%%%%%
\appendix
\section{ Quantum fluctuations of the bound state mode}

In the following two sub-sections we evaluate the amplitude for the
bound state zero mode.
The integration here is done along the closed
contour with the dotted line as depicted in Fig.~3.

\subsection{The two-sphere}

First, we note that the bound state for the minimally coupled case is given 
by
\beq
q_0=\frac{i\nu}{\sigma}~,
\eeq
where the bound state zeta function is defined as
\bea
\tilde\zeta_{\rm bs}(s)& = &
4f_0^2(0)\frac{\mu^{2(s-1)}}{H^{2s}}
\sum_{j=0}^{\infty} (j+1/2) \left( (j+1/2)^2
-\left(\frac{\nu}{\sigma}\right)^2
\right)^{-s} 
\nonumber  \\
&=& 4f_0^2(0)\frac{\mu^{2(s-1)}}{H^{2s}}
\bar \zeta_{\rm bs}(s)~.
\eea
Here, $f_0(z)$ is the normalized mode function of the bound state.
Quite clearly we have a zero mode (by zero mode we mean that the
lowest eigenvalue $\lambda_0$ is a null eigenvalue, i.e.,
$\lambda_0=0$) and in such a case we have to project out this mode to
evaluate the bound state contribution.
However, in general the bound state varies from the top of the mass
gap at $(\nu/\sigma)^2=1/4$ down to $\nu/\sigma=0$, which is for the 
massless
conformally coupled case. In the following we shall focus on a general
bound state mass $\nu/\sigma$ taking care when dealing with the bound state
zero mode. 
 
It is straightforward to evaluate the above $\zeta$-function by employing 
the
binomial expansion method which follows identically to that of Allen
\cite{Allen}, see \cite{CREP} for the case when null
eigenvalues are present. Thus, subtracting out the null eigenvalue we
obtain the following: 
\beq
\bar \zeta_{\rm bs}(s) =\frac{1}{2}\sum_{J=0}^{\infty}
 {\Gamma(s+J)\over J! ~\Gamma(s)}
 \Big[\left(\frac{\nu}{\sigma}\right)^{2J}\zeta_c(s+J)-
 \delta_{\xi\,, 0}\left(\frac{1}{2}\right)^{-2s}\Big]\,, 
\label{bino}
\eeq
with $\zeta_c(s)$ for $S^{2}$ 
defined by
\bea
\zeta_c(s)&=&2 \zeta_{ H} \Big(2s-1,\frac{1}{2}\Big)~,
\eea
which is the zeta function for the bound state mode in the conformally
coupled case, evaluated explicitly in \cite{Naylor:2004ua} in the case
of $d+1=3$.
Essentially the minimally coupled case requires summing from $J=1$ 
instead of $J=0$ in (\ref{bino}), i.e. we have to subtract out the null
eigenvalue.

Similar to the case discussed in \cite{CREP} (section 11.3, Eq.~(11.73),
pp. 80) there is a pole in the above Hurwitz zeta function at $s=1$,
which can be simply inferred from the relation Eq.~(\ref{leurant2d}).
As discussed in
\cite{Naylor:2004ua} a suitable way to deal with the the pole at $s=1$
is to apply the improved zeta function method, described in
\cite{IM,CEZ}, 
which leads to a expression for the amplitude
\bea
\label{modi}
\langle\chi^2(x)\rangle_{\rm bs}
&=&
\lim_{s\to1}
\frac{d}{ds}\left[
(s-1)\,
\zeta_{\rm bs}(s)(x)\right]
\,.  \label{bsamp}
\eea
Note, the above expression agrees with the usual definition when there
is no pole at $s=1$. 

Applying the above equation to our case we obtain 
%the following relation
%\bea
%\lim_{s\to 1}
%\frac{d}{ds} \left[ (s-1) \tilde \zeta_{\rm bs}(s)
%\right] 
%&=&
%2f_0(0)
%\Bigl(
%\mu^{-2}
%\lim_{s\to1}
%\frac{d}{ds}\left[
%\left(\frac{\mu}{H}\right)^{2s}
%(s-1)\,\Big(2\zeta_H(2s-1,\frac{1}{2})-
% \delta_{\xi\,, 0}\left(\frac{1}{2}\right)^{-2s}\Big)\right]\nn
%\nonumber  \\
%&&+\frac{\mu^{2(s-1)}}{H^{2s}}
%\sum_{J=1} {\Gamma(s+J)\over J! ~\Gamma(s)}
% \Big[2\left(\frac{\nu}{\sigma}\right)^{2J}
%   \zeta_H(2s-1+2J,\frac{1}{2})-
% \delta_{\xi\,,0}\left(\frac{1}{2}\right)^{-2s}\Big]\Bigr)\,,
%\nonumber \\
%\eea
%which leads to
\bea
H^2
 \langle
\tilde \chi^2(0)
 \rangle_{bs}
&=&2f_0^2(0)
 \Bigl( 2\ln \left(\frac{\mu}{H}\right)
       -2 \psi(1/2)
        -\delta_{\xi\,, 0} \left(\frac{1}{2}\right)^{-2}
 +    \sum_{J=1}^{\infty} 
 \Big[2\left(\frac{\nu}{\sigma}\right)^{2J}\zeta_H(2J+1,\frac{1}{2})-
 \delta_{\xi\,, 0}\left(\frac{1}{2}\right)^{-2}\Big]
 \Bigr).
\nonumber \\
\eea

Next,~we determine $f_0(0)$.
The normalized bound state solution is
\bea
 f_0^2(z)
=\frac{1}{2\sigma}
 \cosh^{-2\nu}(x) 
 \left(
      \int^{\infty}_0 dy \cosh^{-2\nu}(y)
 \right)^{-1}\,.
\eea 
Thus,
\bea
  f_0^2(0)
= \frac{1}{2\sigma}
 \left(
      \int^{\infty}_0 dy \cosh^{-2\nu}(y) 
 \right)^{-1} \,\label{normbc}.
\eea
Note that for the conformally coupled case $\xi=\xi_c$, $\nu=0$ and 
therefore, the amplitude of the bound state vanishes. This agrees with the 
result found in 
\cite{Naylor:2004ua}, for the thin brane case. In fact numerical plots
of the amplitude for the bound state mode versus the brane thickness show
that the bound state mode is independent of the brane thickness.  

Finally, for the bound state mode, we obtain the normalized amplitude as

\bea
H^2\langle
\tilde \chi^2(0)
 \rangle_{\rm  bs}
&=& \frac{1}{\sigma}
 \Bigl( 2\ln\left(\frac{\mu}{H}\right)
       -2 \psi(1/2)
        -\delta_{\xi\,, 0} \left(\frac{1}{2}\right)^{-2}
 +    \sum_{J=1}^{\infty} 
 \Big[2\left(\frac{\nu}{\sigma}\right)^{2J}\zeta_H  (2J+1,\frac{1}{2})
  - \delta_{\xi\,, 0} \left(\frac{1}{2}\right)^{-2}\Big]
 \Bigr)
\nonumber \\
&\times&
\left(\int^{\infty}_0 dy \cosh^{-2\nu}(y) \right)^{-1}
\,.
\eea
This can now be compared with the result for that of the KK modes.

%%%%%%%%%%%%%%%%%%%%%%%%%%%%%%%%%%%%%%%%%%%%%%%%%%%%%%%%%%%%%
\subsection{The four-sphere}

The zeta function for the bound state can be written
as
\bea
\tilde \zeta_{\rm bs}(s)
&=&
2\frac{1}{H^2}\left(\frac{\mu}{H}\right)^{2(s-1)}
f_0^2(0)
\bar \zeta_{\rm bs}(s)\,, \label{zetabs4d}
\eea
where
\bea
\bar \zeta_{\rm bs}(s)
:=\frac{1}{3}
  \sum_{j=0}^{\infty}
 \frac{(j+3/2)(j+1)(j+2)}
       {[(j+3/2)^2-(\nu/ \sigma)^2]^s}
\eea
is the zeta function for a massive scalar field on $S^4$.
For the $S^4$ geometry, the zeta function for a massless, conformally 
coupled
scalar field is given by
\bea
\zeta_c(z)
=\frac{1}{3}
 \Bigl[
   \zeta_{H}(2s-3,\frac{3}{2})
  -\frac{1}{4}
   \zeta_{H}(2s-1,\frac{3}{2}) 
 \Bigr]\,.
\eea
Thus, the dS zeta function for a general mass can be written as a summation
over the massless conformal zeta functions (by employing the binomial 
expansion)
\bea
\bar \zeta_{\rm bs}(s)
&=&
\sum_{J=0}^{\infty}
 {\Gamma(s+J)\over J! ~\Gamma(s)}
 \Big[\left(\frac{\nu}{\sigma}\right)^{2J}\zeta_c(s+J)-
 \delta_{\xi\,,0}\left(\frac{3}{2}\right)^{-2s}\Big] 
\nonumber \\
&=&\sum_{J=0}^{\infty}
 {\Gamma(s+J)\over J! ~\Gamma(s)}
 \Big[\frac{1}{3}
     \left(\frac{\nu}{\sigma}\right)^{2J}
   \Bigl\{
      \zeta_H(2s-3+2J,\frac{3}{2})
     -\frac{1}{4} \zeta_H(2s-1+2J,\frac{3}{2})
   \Bigr\}
 - \delta_{\xi\,,0}\left(\frac{3}{2}\right)^{-2s}\Big] 
\nonumber \\
&=&
 \frac{1}{3}
  \Bigl[
       \zeta_{H}(2s-3,\frac{3}{2})
    + s\left(\frac{\nu}{\sigma}\right)^2
       \zeta_{H}(2s-1,\frac{3}{2}) 
-\frac{1}{4}\zeta_H(2s-1,\frac{3}{2})
-\frac{1}{4}s\left(\frac{\nu}{\sigma}\right)^2
    \zeta_H(2s+1,\frac{3}{2})
\nonumber  \\
& +& \sum_{J=2}^{\infty}
 {\Gamma(s+J)\over J! ~\Gamma(s)}
     \left(\frac{\nu}{\sigma}\right)^{2J}
      \zeta_H(2s-3+2J,\frac{3}{2})
  -\frac{1}{4}  
    \sum_{J=2}^{\infty}
 {\Gamma(s+J)\over J! ~\Gamma(s)}
     \left(\frac{\nu}{\sigma}\right)^{2J}
      \zeta_H(2s-1+2J,\frac{3}{2})
    \Bigr]
\nonumber\\
&-&\sum_{J=0}^{\infty}
    \frac{\Gamma(s+J)}{J!\Gamma(s)}
 \delta_{\xi\,,0}\left(\frac{3}{2}\right)^{-2s}\,.
\eea

Now, we can evaluate the% zeta function for the bound 
%state, Eq.~(\ref{zetabs4d}), and
~squared amplitude of the bound state from Eq.~(\ref{bsamp}).
%From Eq.~(\ref{leurant4d}), we obtain
%\bea
%&&\lim_{s\to1}
%\frac{d}{ds}\left[(s-1)\tilde \zeta_{\rm bs}(s)(x)\right]
%\nonumber  \\
%&=&
%\frac{ f_0^2(z)}{H^2}
%\Bigl\{
%    \left(
%            -\frac{1}{6}
%           +\frac{2}{3}
%          \left(\frac{\nu}{\sigma}\right)^2
%    \right)
%    \ln(\frac{\mu}{H})     
%+ \frac{2}{3}\zeta_{H}(-1,\frac{3}{2})
%     +\frac{1}{6}\psi(3/2)
%-\left(\frac{\nu}{\sigma}\right)^2
%\left(
%   \frac{2}{3}\psi(3/2)
%   +\frac{1}{6}\zeta_H (3,\frac{3}{2})
%\right)
%\nonumber \\  
% &+&\frac{2}{3}
%      \sum_{J=2}^{\infty}\left(\frac{\nu}{\sigma}\right)^{2J}
% \left(  
% \zeta_H(2J-1,\frac{3}{2}) 
% -\frac{1}{4}  \zeta_H(2J+1,\frac{3}{2}) 
% \right)
%     -2\sum_{J=0}^{\infty}\delta_{\xi\,, 0}
%       \left(\frac{3}{2}\right)^{-2} 
%\Bigr\}\,.
%\eea
The normalization of the bulk mode is the same as the $d=2$ case and
at $z=0$ we obtain Eq.~(\ref{normbc}) with $\nu$ for $d=4$,  
%\bea
%f_0^2(0)=\frac{1}{2\sigma}
%           \left(
%                \int^{\infty}_0 dy \cosh^{-2\nu}(y)
%           \right)^{-1}\,,
%\eea
%where
\bea
\nu=\frac{1}{2}\left(
       \sqrt{1+(3-16\xi)(3\sigma+2\sigma^2)}-1
        \right)\,.
\eea
The resultant bound state amplitude is
\bea
H^2
\langle  \tilde \chi^2(0)\rangle_{\rm bs}
&=&
\frac{1}{2\sigma}
           \left(
                \int^{\infty}_0 dy \cosh^{-2\nu}(y)
           \right)^{-1}
\nonumber  \\
&\times&
\Bigl\{
  \left(
            -\frac{1}{6}
           +\frac{2}{3}
          \left(\frac{\nu}{\sigma}\right)^2
    \right)
    \ln\left(\frac{\mu}{H}\right)     
+ \frac{2}{3}\zeta_{H}(-1,\frac{3}{2})
     +\frac{1}{6}\psi(3/2)
-\left(\frac{\nu}{\sigma}\right)^2
\left(
   -\frac{1}{3}
  + \frac{2}{3}\psi(3/2)
   +\frac{1}{6}\zeta_H (3,\frac{3}{2})
\right)
\nonumber \\  
 &+&\frac{2}{3}
      \sum_{J=2}^{\infty}\left(\frac{\nu}{\sigma}\right)^{2J}
 \left(  
 \zeta_H(2J-1,\frac{3}{2}) 
 -\frac{1}{4}  \zeta_H(2J+1,\frac{3}{2}) 
 \right)
     -2\sum_{J=0}^{\infty}\delta_{\xi\,, 0}
       \left(\frac{3}{2}\right)^{-2} 
\Bigr\}\,.
\eea

%%%%%%%%%%%%%%%%%%%%%%%%%%%%%%%%%%%%%%%%%%%%%%%%%%%%%%
\section{Classical Stability against tensor and scalar perturbations}

Here, we briefly discuss the stability of the thick brane model
both against tensor and scalar perturbations for general $d$-dimensions.

%%%%%%%%%%%%%%%%%%%%%%%%%%%%%%%%%%%%%%%%%%%%%%%%%%%%%%%%%%
\subsection{Tensor perturbations}

We first discuss the tensor perturbations about the domain wall background.
Here we shall assume a Randall-Sundrum~(RS) type gauge \cite{Randall:1999vf} in which
the components of the extra-dimension are zero, i.e.,
\begin{eqnarray}
ds^2=
b^2(z)
\left[dz^2
  +\left(
    \gamma_{\mu\nu}+h_{\mu\nu}
  \right)
dx^{\mu}dx^{\nu}
\right]\,,
\end{eqnarray}
where $h_{\mu\nu}$ satisfies the usual transverse-traceless gauge about
the background dS metric;
$D^{\mu}h_{\mu\nu}=h^{\mu}{}_{\mu}=0$, where
$D^{\mu}$ is the covariant derivative associated with $\gamma_{\mu\nu}$.

In this case, the perturbation is separable and we obtain the equation
of motion in the bulk direction,
which can be written in the standard quantum mechanical
form as
\begin{eqnarray}
\Bigl[-\frac{d^2}{dz^2} + V_T(z)\Bigr]\psi(z)= m^2\psi(z)\,,
\end{eqnarray}
where $\psi(z) \propto b(z)^{-(d-1)/2}h_{\mu\nu}$ and
\begin{eqnarray}
V_T (z) =\frac{(d-1)^2}{4}H^2
        -\frac{d-1}{4}H^2\frac{d-1+\frac{2}{\sigma}}{\cosh^2(H z/\sigma)} \,.
\label{tensorpotential}
\end{eqnarray}

The thin wall limit can be obtained from the limit $\sigma \to 0$, which 
leads to a system composed of a thin dS brane embedded in a flat Minkowski 
bulk (for quantum fluctuations in such a model, see e.g., 
\cite{Pujolas:2004uj}). 
The potential for the tensor perturbations in the thin wall limit 
is then
\begin{eqnarray}
V_T(z)=\frac{(d-1)^2}{4}H^2-(d-1)H\delta(z)\,,
\end{eqnarray}
where we used for $\sigma \to 0$
\begin{eqnarray}
 \frac{1}{2\sigma\cosh^2(x/\sigma )}\to \delta(x)\,. 
\end{eqnarray}
In this limit the solution
for the tensor perturbations reduces to the standard exponential form. 

The general solution can be decomposed into a zero mode with mass
$m=0$, which may realize four-dimensional gravity on the 
brane, and a continuous spectrum of Kaluza-Klein (KK) modes with $m>3/2$ 
(in the
five-dimensional case).  
Thus, the model is classically stable against the tensor perturbations.

\subsection{Scalar perturbations}

Next, we discuss the stability of the model against scalar perturbations. 
We consider a scalar metric perturbation of the form
\bea
ds^2= b(z)^2 
    \Bigl[
          \Bigl(1+2A\Bigr)  dz^2
        + 2 D_{\mu} B dx^{\mu} dz
        +\Bigl(\gamma_{\mu\nu}
               \Bigl( 1+2{\cal R}
               \Bigr)  
        +2  D_{\mu}D_{\nu}E
          \Bigr)dx^{\mu}dx^{\nu}
     \Bigr]
\eea
and also a perturbation of the field $\phi(z)\to \phi(z)+\delta
\phi(x)$, which supports the domain wall. 

In the bulk longitudinal gauge, $B=E=0$, the perturbed Einstein equations 
can be written as follows:
\begin{eqnarray}
d(d-1)\frac{b'}{b}{\cal R} '
      +(d-1)\Box{\cal R}
      +d(d-1)H^2{\cal R}
      -d(d-1)\Bigl(\frac{b'}{b}\Bigr)^2A    
%\nonumber \\ 
%&&\hspace{5cm}
&=&\phi' \delta\phi'
          -A\phi'^2
          -b^2 \frac{\partial V}{\partial \phi}\delta\phi\,,
\nonumber\\
-(d-1)D_{\mu}
      \Bigl(
          {\cal R}'
          -\frac{b'}{b}A
      \Bigr)  
   &=&  \phi' D_{\mu}\delta\phi \,,
\nn
      (d-1){\cal R}''
          +(d-1)^2\frac{b'}{b}{\cal R}'
          +(d-2)\Box {\cal R}+(d-2)(d-1)H^2{\cal R}  
             -(d-1)\frac{b'}{b} A'
          &-&2(d-1)\frac{b''}{b}A   %%%%%%
          +\Box  A
\nonumber \\ 
%\nonumber \\
%&&
&=&       
 A \phi'^2
                -\phi'\delta\phi'
               -b^2 \frac{\partial V}{\partial \phi}\delta\phi\,,
\nonumber \\
D^{\alpha}D_{\beta}\Bigl((d-2){\cal R}+A\Bigr)&=&0\,. 
\end{eqnarray}
The perturbed equation of motion of the scalar field is found to be
\begin{eqnarray}
            \delta\phi''
           +(d-1)\frac{b'}{b}\delta \phi'
            +\Box \delta\phi             
            -2A
           \Bigl(
                \phi''
               +(d-1)\frac{b'}{b}\phi'
           \Bigr)
            + \Bigl(
               -A'
               +d{\cal R}'
               \Bigr)\phi'
               -b^2\frac{\partial^2 V}{\partial \phi^2}
             \delta\phi
          =0\,. 
\label{pertfield}
\end{eqnarray}

Next, we derive the evolution equation for the curvature perturbation 
${\cal R}$.
By defining
\begin{eqnarray}
 \Psi= {\cal R} \Bigl(\frac{\phi'^2}{b^{d-1}}\Bigr)^{-1/2}
\end{eqnarray}
the equation for the curvature perturbations can be reduced to the form
\bea
-\Psi'' +V_{S}  \Psi=  \Box_d \Psi
\eea
with potential
\begin{eqnarray}
V_S= 
      \frac{d^2+4d-13}{4}\Bigl(\frac{b'}{b}\Bigr)^2
     -\frac{3d-7}{2}\frac{b''}{b}
     +(d-3)\frac{b'}{b}\frac{\phi''}{\phi'}
     -\frac{\phi'''}{\phi'}
     +2\Bigl(\frac{\phi''}{\phi'}\Bigr)^2
     -2(d-1)H^2\,.
\end{eqnarray} 

For the dS thick brane case, which is considered in this article, we obtain
\begin{eqnarray}
V_S= \frac{\beta^2}{4\cosh^2(\beta  z)}
\Bigl\{
  2\bigl[2+(3d-7)\sigma-(4d-4)\sigma^2\bigr]
+\Bigl[4+4(d-3)\sigma+(d^2-10d+9)\sigma^2\bigr]
 \sinh^2(\beta z)
\Bigr\}\,.
\end{eqnarray}
Thus, it is simple to see that at least both for the cases of
interest, $d=2$ and $d=4$, $V_S>0$ and therefore, the model is always
stable against scalar perturbations. The $d=4$ case was originally
derived in \cite{Wang:2002pk}.
 
%%%%%%%%%%%%%%%%%%%%%%%%%%%%%%%%%%%%%%%%%%%%%%%%%%%%%%%%%%%%%%

%%%%%%%%%%%%%%%%%%%%%%%%%%%%%%%%%%%%%%%%%%%%%%%%%%%%%%%%%%%%%%%%%%%%%%%

\end{document}